# A Family of Finite-Temperature Electronic Phase Transitions in Graphene Multilayers

Youngwoo Nam[1,2], Dong-Keun Ki[1,3], David Soler-Delgado[1], and Alberto F. Morpurgo[1]*

[1]Department of Quantum Matter Physics (DQMP) and Group of Applied Physics (GAP), University of Geneva, 24 Quai Ernest-Ansermet, CH1211 Genéve 4, Switzerland.

[2]Department of Physics, Gyeongsang National University, Jinju-daero 501, Jinju-si, South Korea

[3]Department of Physics, The University of Hong Kong, Hong Kong, China

* Correspondence to: alberto.morpurgo@unige.ch

**Abstract:**

Suspended Bernal-stacked graphene multilayers up to an unexpectedly large thickness exhibit a broken-symmetry ground state, whose origin remains to be understood. Here we show that a finite-temperature second order phase transition occurs in multilayers whose critical temperature $T_c$ increases from 12 K in bilayers to 100 K in heptalayers. A comparison of the data to a phenomenological model inspired by a mean field approach suggests that the transition is associated with the appearance of a self-consistent valley- and spin-dependent staggered potential changing sign from one layer to the next, appearing at $T_c$ and increasing upon cooling. The systematic evolution with thickness of several measured quantities imposes constraints on any microscopic theory aiming to analyze the nature of electronic correlations in this system.

**Main text:**

Clean two-dimensional conductors in the presence of a large perpendicular magnetic field are strongly correlated systems. Their ground states are determined by Coulomb repulsion between electrons and are characterized by broken symmetries and non-trivial topological invariants that depend sensitively on electron density ($n$) and applied field ($B$) (*1-4*). Accordingly, a series of quantum phase transitions occurs upon varying $n$ and $B$, which manifest themselves in the very rich phenomenology ubiquitously observed in magneto-transport measurements (*5-13*). These considerations hold true irrespective of the specific two-dimensional (2D) conductor considered, because they rely almost exclusively on the formation of Landau levels that quench the electron kinetic energy, and allow electron-electron (*e-e*) interactions to dominate. In the absence of Landau levels, however, *e-e* interactions play a much less prominent role.

Multiple recent experiments indicate that in graphene multilayers this is not the case (*14-19*). A gapped insulating state at $B = 0$, first reported in Bernal-stacked bilayers (*14-17*), has been recently observed in all even Bernal-stacked multilayers up to octalayer graphene (8LG) (*18, 19*). In contrast, odd Bernal-stacked multilayers (so far mono and trilayer have been studied) remain conducting at low temperature (*20-23*). These findings defy common expectations, namely that the behavior of graphene multilayers should approach that of graphite as thickness is increased. No direct information about the nature of the insulating state in even multilayers could be obtained so far, because the phenomenon is only observed in suspended graphene devices of the highest quality (*14-23*), which makes measurements other than transport challenging. Because, upon cooling, the resistance of even multilayers increases without showing abrupt changes at any specific temperature, it's even unclear whether the insulating state results from a quantum phase transition (with a gap present at all temperatures), or if a phase transition occurs at a critical temperature $T_c$ with the gap vanishing for $T > T_c$ (*16, 17, 24-28*). To address these issues, here we study ultra-clean, suspended Bernal-stacked multilayers of graphene near charge neutrality and show that at $B = 0$ these systems unambiguously exhibit an *e-e* driven finite-temperature phase transition to a broken symmetry state at a critical temperature $T_c$ that depends on thickness.

To draw these conclusions, we focused on the density of electrons present in the conduction band of charge-neutral multilayers, $n_{th}(T)$, whose temperature dependence is determined by the low-energy density of states (DOS). At sufficiently low $T$, $n_{th}(T)$ should show an exponential increase in the presence of gap –given that at charge neutrality electrons in the conduction band are thermally activated from the valence band– or stay constant if an overlap between valence and conduction band is present. If the multilayers are zero-band gap semiconductors, $n_{th}(T)$ is expected to increase with increasing $T$, with a specific dependence determined entirely by the low-energy DOS. Quantitative information can be obtained by comparing experimental data to the dependence of $n_{th}(T)$ calculated from a chosen theoretical expression for the low-energy DOS, as we illustrate for the case of bilayers. Under the assumption that only the nearest neighbor in-plane ($\gamma_0$) and out-of-plane ($\gamma_1$) hopping terms are relevant (the so-called minimal tight-binding model), the low-energy quadratic dispersion relation of electrons in bilayer graphene (Fig. 1A) leads to an energy-independent DOS (Fig. 1B). The resulting density $n_{th}(T)$ of electrons in the conduction band then increases linearly with temperature (Fig. 1C). However, if a gap $\Delta$ opens (Fig. 1D), the DOS is modified (Fig. 1E), and so is the temperature dependence of $n_{th}(T)$. Figure 1F shows $n_{th}(T)$ expected for a gap $\Delta$ exhibiting a mean-field temperature dependence and vanishing at $T_c$ (see the inset and (*29*)). A transition becomes visible at $T = T_c$, below which $n_{th}(T)$ is pronouncedly suppressed as compared to the non-interacting case.

What makes these considerations useful is that the extremely small charge inhomogeneity present in suspended graphene multilayers enables the density $n_{th}(T)$ of thermally excited electrons to be determined experimentally over a broad range of temperatures. $n_{th}(T)$ can be extracted from the density dependence of the conductance $G(n)$ near charge neutrality. To understand why and how, it is useful to look at the double logarithmic plot of $G(n)$ (inset of Fig. 1G). We see that there is a range of $n$ over which $log(G)$ is constant, and that $log(G)$ starts increasing significantly only when $n$ becomes larger than a threshold (that we denote $n^*(T)$; $n^*(T)$ increases at higher $T$). The physical reason is clear: the square conductance remains virtually unchanged as long as the density $n$ of electrons accumulated by the gate voltage is much smaller than the density of electrons $n_{th}(T)$ already present owing to thermal activation from the valence band. Finding that the square conductance increases significantly only when $n$ exceeds $n^*(T)$ therefore implies that $n_{th}(T) \sim n^*(T)$, as discussed previously for mono and bilayer graphene (*21, 30-32*). In practice, the exact value of $n^*(T)$ depends on the criterion used for its definition: here we define $n^*(T)$ as the value of $n$ for which the conductivity $\sigma(n,T) = 1.67\sigma(n=0,T)$, a relation that we can validate for bilayers and that is approximately correct for thicker multilayers (*29*). We emphasize, however, that none of our key conclusions depends on the precise criterion used (*29*).

The temperature dependence of $n^*(T)$ extracted from measurements on four different bilayer devices is shown in Figure 1G. For all devices, a critical temperature $T_c \cong 12$ K is found, below which $n^*(T)$ is suppressed as compared to the gapless case. The red continuous curve, which represent $n_{th}(T)$ expected for a gap having a mean-field temperature dependence and $\Delta_0 = \Delta(T = 0) \cong 1.9$ meV, reproduces the data very satisfactorily. The presence of a clear transition below which a finite $\Delta$ appears, the value of $T_c$, and the shape of $n^*(T)$ are very robust against all aspects of the data analysis. They can be extracted directly from the data without any assumption about the DOS, and do not depend on the criterion used to extract $n^*(T)$ (only the quantitative determination of $\Delta$ does require the DOS to be specified, see (*29*)). Remarkably, the same value for the critical temperature is found in all the different devices measured, despite the fact that the absolute value of the low-temperature resistance –and how pronounced the insulating state is at the lowest temperature of our measurements– exhibit much larger sample-to-sample fluctuations (because of these fluctuations, even the occurrence of an insulating state driven by *e-e* interaction has been questioned in some past experimental work (*14-17, 30, 33*)). The reason for this excellent reproducibility is that our measurements effectively probe the DOS averaged over the entire device area, whereas in the insulating state the absolute value of the resistance is strongly affected by any percolating conducting path (e.g. at edges (*34, 35*)) that occupies a negligible fraction of the total area, giving a negligible contribution to the DOS. The highly reproducible behavior of $n^*(T)$ allows us to establish unambiguously the occurrence of an insulating, gapped state near charge neutrality, which is entered through a second-order phase transition at $T_c = 12$ K.

The same strategy outlined for bilayers can be applied to thicker even multilayers, in which multiple conduction and valence bands are present (*29, 36-39*). Recent work (specifically, the quantization of the Hall effect at low magnetic field and magneto-Raman experiments) has provided evidence that in suspended devices at low energy these bands are well described by including only nearest neighbor in plane ($\gamma_0$) and out of plane ($\gamma_1$) hopping terms (*18, 19, 40, 41*). According to this description, all conduction and valence bands in even multilayers are quadratic at low energy (with different effective masses), so that the DOS in the non-interacting case is again energy-independent (see (*29*)). The same argument used for bilayers implies that $n_{th}(T)$ increases linearly with temperature in the absence of interactions, and that the opening of a gap causes a suppression for $T < T_c$. Figure 2, A and C illustrates the results of experiments performed on tetralayer (4LG) and hexalayer (6LG) graphene devices: indeed $n^*(T)$ depends linearly on temperature at sufficiently high temperature, but below a critical temperature $T_c$ (= 38 K for 4LG and 90 K for 6LG) $n^*(T)$ is suppressed, just as for bilayers (Fig. 1, F and G). For 4LG and 6LG, the data are excellently reproduced by assuming that a gap $\Delta$ with a mean-field temperature dependence opens simultaneously at $T_c$ on all quadratic bands (red line; $\Delta_0 \cong 5.2$ meV for 4LG and $\Delta_0 \cong 13$ meV for 6LG). As in the bilayer case, the temperature dependence of the resistance at charge neutrality (Figure 2, B and D) demonstrates the insulating nature of thicker even multilayers, but provides no specific feature allowing the identification of a critical temperature.

The analysis of the measured temperature dependence of $n^*(T)$ and the comparison with the calculated expression for $n_{th}(T)$ can be performed – and interpreted in the same way– also for odd Bernal multilayers. Within the minimal tight-binding model mentioned above, odd multilayers contain multiple conduction and valence bands, all having a quadratic dispersion, except for one that is a linear Dirac band (*29, 36-39*). The DOS associated to the linear band vanishes at charge neutrality, and contributes at most a few percent of the total DOS in the temperature and density range relevant for our work (*29*). It is therefore negligible in practice, so that the same considerations made for even multilayers hold true for odd ones. In Fig. 3, data measured on odd

Bernal-stacked multilayers show that this is indeed the case for both trilayer (3LG, Fig. 3A) and heptalayer graphene (7LG, Fig 3C): $n^*(T)$ increases linearly with temperature at high temperature, exhibiting a pronounced suppression for $T < T_c$ (with $T_c = 33$ K for 3LG and 100 K for 7LG), in complete analogy to the case of even multilayers. Like for even multilayers, the data agree quantitatively with the behavior expected if a gap $\Delta$ opens simultaneously at $T_c$ in all quadratic bands, with a mean-field temperature dependence and $T = 0$ values of $\Delta_0 \cong 5$ meV for 3LG and $\Delta_0 \cong 13$ meV for 7LG. For odd multilayers the low-temperature conductivity remains finite, $\sigma \sim e^2/h$, indicating that the linear band does not gap out (Fig. 3, B and D).

In summary, we find that a phase transition occurs in all the Bernal-stacked multilayer investigated, irrespective of whether they are even or odd, with only the even ones becoming insulating at low temperature. Data for multilayers of all thicknesses are shown in Figure 4A– each with the corresponding fit to the calculated temperature dependence of the density of thermally activated electrons $n_{th}(T)$. Remarkably, if $n^*(T)/n^*(T_c)$ is plotted versus $T/T_c$ all data collapse on top of each other (Fig. 4A, inset). $T_c$ increases linearly with increasing thickness (Fig. 4B) and so does the gap $\Delta_0$, that is proportional to $T_c$ (Fig. 4C): the best linear fit –continuous line– is close to the expected mean-field value, $\Delta_0/k_BT_c = 1.76$ –dashed line (*29*).

The experimental findings consistently provide substantial evidence about the microscopic nature of the broken-symmetry state, namely that the effect of interactions is well described by a (valley- and spin-dependent) mean-field staggered potential changing sign from one layer to the next, acting as an order parameter (*18, 19*). A scenario based on the minimal tight-binding model augmented with a self-consistent staggered potential explains *i)* that at low temperature even multilayers become fully insulating whereas in odd ones a finite conductivity of ~ $e^2/h$ persists (see Fig. 2, B and D and Fig. 3, B and D); *ii)* that from 1LG to 8LG the first quantum Hall plateau appears systematically at filling factor $\nu = 2N$ with $\sigma_{xy} = 2Ne^2/h$, where $N$ is the multilayer thickness (see (*18, 19*)); *iii)* the occurrence of the transition in odd multilayers; *iv)* that the gap in the different quadratic bands has the same magnitude; *v)* the simultaneous opening of the gap in all bands at the same critical temperature (indeed, inasmuch *iv)* and *v)* are concerned, a generic mechanism could gap each band independently from the others, resulting in multiple transitions with different values of $T_c$ and $\Delta$ (*42, 43*)). Note that in the scenario that we propose –with the staggered potential acting as order parameter– the symmetry broken is discrete, which is why the occurrence of finite-temperature phase transitions in Bernal multilayers does not conflict with the Mermin-Wagner theorem (*44*), despite all multilayers being effectively two-dimensional.

Despite the success of the proposed model, in the absence of a comprehensive theoretical analysis important questions remain. A key one is the validity of the minimal tight-binding model because the vast existing literature on graphite suggests that more tight-binding parameters should be included. A recent theoretical analysis of *e-e* interaction at the Hartree-Fock level (*43*) shows that adding hopping terms present in graphite –and most notably the parameter $\gamma_2$, responsible for the semi-metallic behavior of graphite ($\gamma_2$ ~ -20 meV usually)– would make the systematic behavior observed in multilayers impossible to reproduce theoretically. At the simplest level the same holds true for other hopping parameters, e.g., $\gamma_3$, responsible for trigonal warping. Why some hopping terms that must be included to describe graphite do not appear in thin multilayers requires an explanation. A key difference may be that our experiments probe charge neutral multilayers (i.e., $n \sim 0$), whereas in graphite approximately $n \sim$ 3-4 $10^{11}$ electrons/$cm^2$ are present in each individual layer (*45*). Indeed it has been established theoretically and experimentally that, as $n$ approaches charge neutrality, large renormalization effects drastically change the hopping parameters in graphene: in monolayers, for instance, the Fermi velocity (and hence $\gamma_0$) is predicted to diverge as $n \rightarrow 0$ (*22, 46-48*). Although a thorough analysis of this renormalization process for all hopping parameters is lacking, what is known at this stage implies that there is no compelling reason why the hopping terms used to describe graphite and charge neutral thin multilayers should be the same.

Irrespective of these details, what is most remarkable in the experimental findings is the occurrence of a phase transition with increasingly large $T_c$ in graphene multilayers that, until now, may have been expected to behave as bulk graphite. Why precisely $T_c$ increases with increasing thickness or at which thickness this trend breaks down, and the behavior of bulk graphite is recovered, remains to be understood. At this stage, a thorough microscopic theoretical analysis is required not only to understand in detail the properties of this family of electronic systems but also to disclose its full potential to reveal subtle phenomena emerging from the physics of correlated electrons.

**Acknowledgments:**

We gratefully acknowledge A. Ferreira for continued technical support of the experiments and T. Giamarchi for useful discussions. **Funding:** Financial support from the Swiss National Science Foundation, the NCCR QSIT, and the EU Graphene Flagship Project are also gratefully acknowledged.

**Author contributions:** YN, DKK, and DSD fabricated the devices and performed the electrical measurements. YN and DKK analyzed the data under the supervision of AFM. All authors discussed experimental results and their interpretation. AFM wrote the manuscript with input from all authors.

**Competing interests:** There are no competing financial interests.

**Data and materials availability:** Data set used in this paper are available at Harvard Dataverse (https://dataverse.harvard.edu/dataset.xhtml?persistentId=doi:10.7910/DVN/9QKJU2).

**Supplementary Materials:**

Materials and Methods

Supplementary Text

Figures S1-S11

References *(50)-(61)*

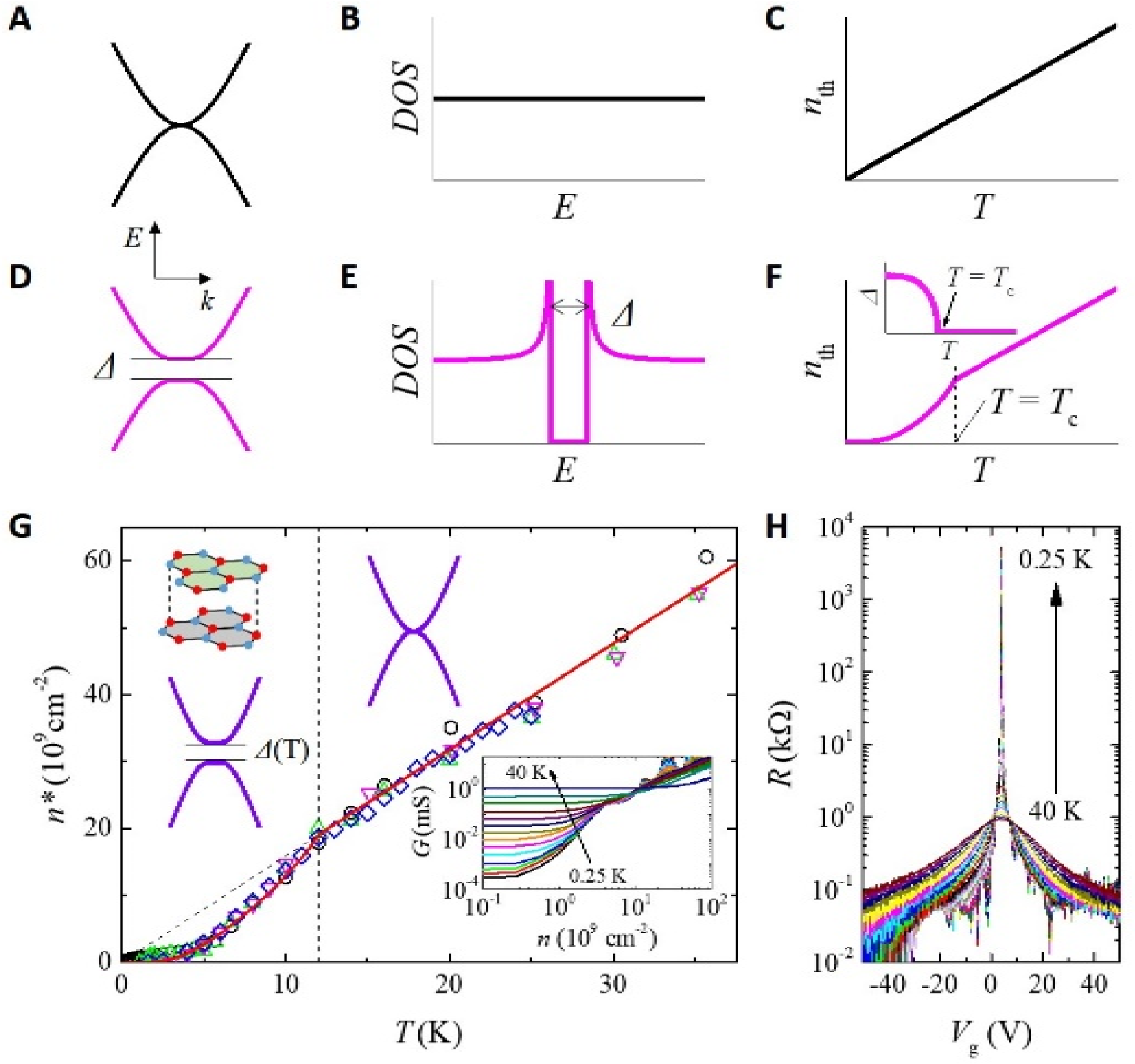

**Fig. 1. Electronic phase transition in bilayer graphene (2LG). (A)** In the absence of interactions, Bernal-stacked graphene bilayers are zero-gap semiconductors with valence and conduction bands dispersing quadratically at low energy, resulting in **(B)** a constant DOS. **(C)** The density of electrons thermally excited from the conduction to the valence band, $n_{th}(T)$, increases linearly with temperature **(D)** The opening of a gap ($\Delta$) thanks to interactions leads to **(E)** a modified DOS and causes **(F)** a suppression of $n_{th}(T)$ (inset: $\Delta(T)$ expected from a mean-field description, used to calculate $n_{th}(T)$). Each symbol in (**G**) represents $n^*(T)$ (~ $n_{th}(T)$ as discussed in the main text and (*29*)) measured on a distinct bilayer device, showing a transition at $T_c$ = 12 K in all cases. Above $T_c$, the value of the slope of $n^*(T)$ matches the one expected from the known DOS of 2LG (i.e.,

effective mass). The red line is a fit to the expression for $n_{th}(T)$ calculated with a mean-field temperature dependence of $\Delta(T)$ (29). Bottom inset: the double logarithmic plot of $G(n)$ measured at different values of $T$ (from bottom to top: 0.27, 0.33, 0.43, 0.53, 0.78, 1.1, 1.4, 2.0 to 5.0 K in 1 K step, 6.7 K, 10 K, and 40 K), from which $n^*(T)$ is extracted. **(H)** The gate-voltage dependence of the resistance of this same device for the same range of temperatures (from top to bottom: 0.25, 0.27, 0.29, 0.33, 0.38, 0.43, 0.48, 0.53, 0.62, 0.78, 0.94, 1.1, 1.2, 1.4, 1.6, 1.7, 2.0 to 5.0 K in 0.5 K step, 5.0, 5.7, 6.7 K, 8.0 to 16 K in 2 K step, and 20 to 40 K in 5 K step) as in (G) exhibits a pronounced insulating behavior around charge neutrality, but no indications of a phase transition.

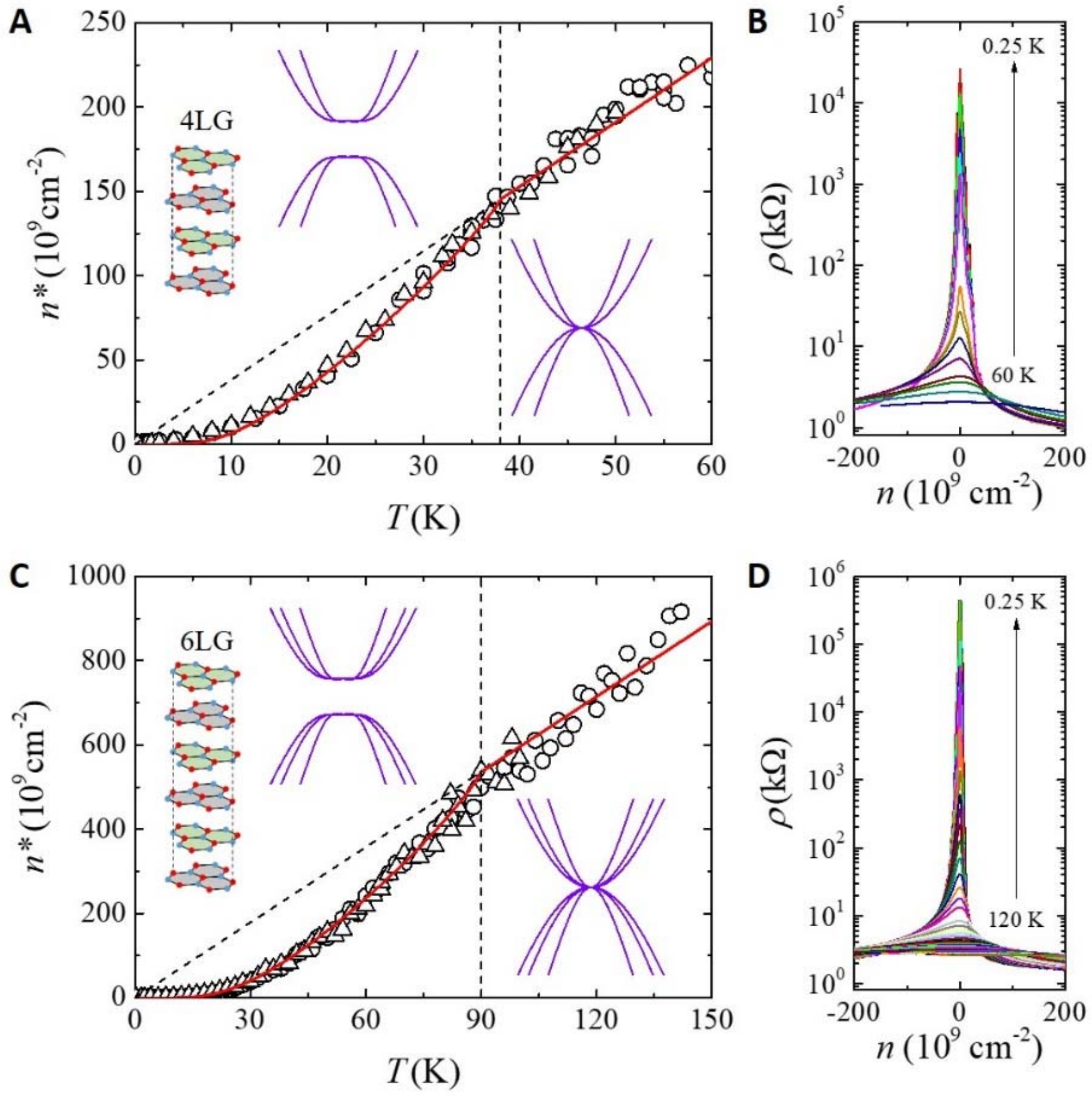

**Fig. 2. Electronic phase transition in even Bernal-stacked multilayers.** Shown are the data on **(A, B)** tetralayer (4LG) and **(C, D)** hexalayer graphene (6LG). (A) and (C) show that at sufficiently high temperature the measured $n^*(T)$ depends linearly on $T$. Below the critical temperature $T_c$ (= 38 K for 4LG and 90 K for 6LG) $n^*(T)$ is suppressed as found in bilayers (Fig. 1). The red continuous curves in (A) and (C), which correspond to the dependence of $n_{th}(T)$ expected for a gap having a mean-field temperature dependence, reproduce the data. In (A) and (C), the circles and triangles represent data obtained from different devices, demonstrating the excellent reproducibility of our observations. (**B**, **D**) Charge carrier density dependence of the resistivity at different temperatures (from top to bottom: in (B) 0.25 to 1.75 K in 0.3 K step and 10 to 55 K in 5 K step, whereas in (D) 0.25 K, 0.4 to 1.7 K in 0.3 K step, 2.5 K, and 4.0 to 100 K in 2 K step), showing an insulating state around charge neutrality at low temperature. The asymmetry visible for positive and negative values of $n$ is caused by the formation of a *pn* junction at the contacts, as typically observed in two-terminal suspended graphene devices. In all multilayers for which four-terminal devices (*49*) could be realized the resistivity is nearly perfectly symmetric upon changing the sign of $n$ (see, e.g., Figs. 1H and 3B).

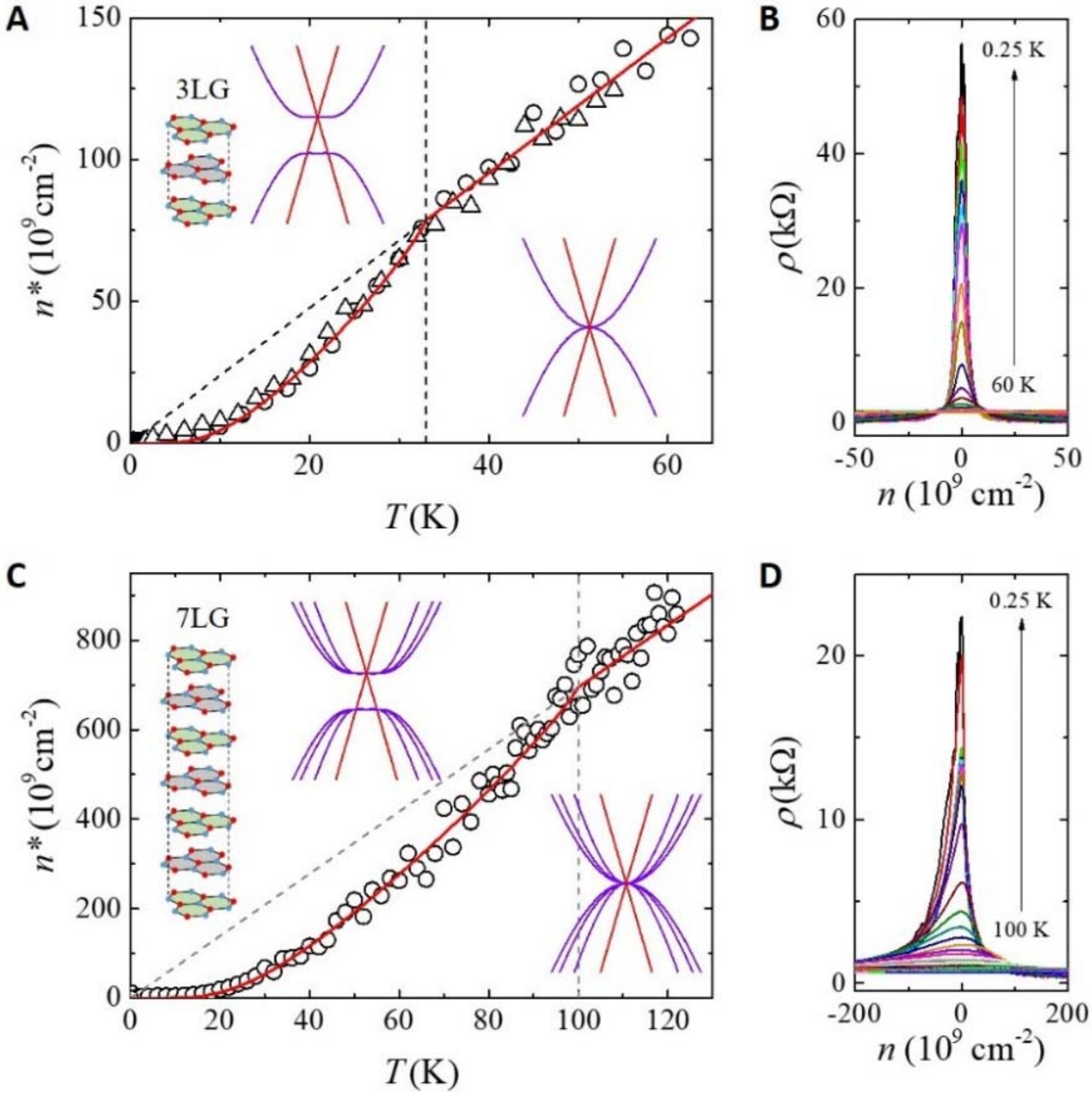

**Fig. 3. Electronic phase transition in odd Bernal-stacked multilayers.** Shown are the data on (**A, B**) trilayer (3LG) and (**C, D**) heptalayer graphene (7LG). (A) and (C) show that the temperature dependence of $n^*(T)$ is linear at sufficiently high $T$, just as observed for even layers, because the DOS associated to the Dirac linear band present in odd layers is negligible in the energy range probed by the experiments. For temperatures below $T_c$ (=33 K for 3LG and 100 K for 7LG) $n^*(T)$ is suppressed, demonstrating that all the quadratic bands gap out for $T < T_c$. The red continuous curves were calculated assuming a mean-field gap. In (A), the circles and the triangles represent data obtained from two different devices. (B) and (D) show charge carrier density dependence of the resistivity at different temperatures (from top to bottom: in (B) 0.25 to 1.5 K in 0.25 K step,

2.5 K, 4.0 to 8.0 K in 2 K step, and 10 to 60 K in 2.5 K step, whereas in (D) 0.24 K, 1.64 K, 2.5 K, and 4.0 to 100 K in 2 K step), showing a finite low-temperature conductivity of ~ $e^2/h$ around charge neutrality, caused by the presence of the linear band that remains un-gapped. The 3LG data in (B), measured on a four-terminal device, are nearly perfectly symmetric in *n*; for the 7LG device in (D) the asymmetry stems from the two-terminal device configuration.

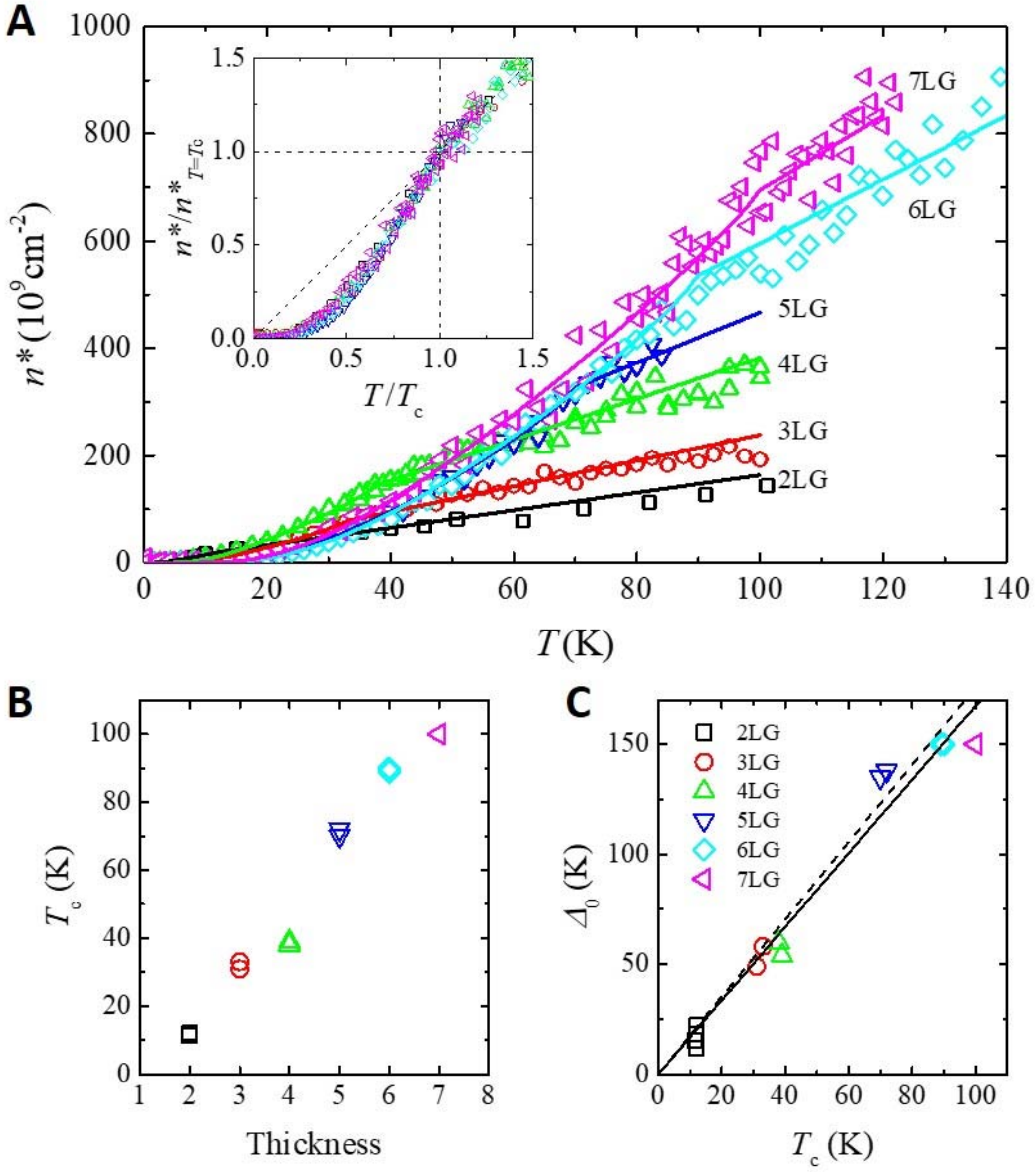

**Fig. 4. Evolution with thickness of the phase transition to the broken-symmetry state.** (**A**) $n^*$ versus temperature for Bernal-stacked multilayers of all thicknesses, from bi to heptalayer graphene. The continuous lines of the corresponding color are obtained by fitting the theoretical expression for the temperature dependence of the density of electron thermally excited from the valence to the conduction band, $n_{th}(T)$, calculated using a mean-field temperature dependence for the gap. Inset: for normalized quantities –i.e., $n^*(T)/n^*(T_c)$ plotted versus $T/T_c$– all curves collapse

on top of each other. For every thickness, only one critical temperature is observed and only one value of the gap is needed to reproduce the data, indicating that a gap opens simultaneously in all quadratic bands. (**B**) $T_c$ increases linearly with increasing thickness and the gap $\Delta_0$ is proportional to $T_c$ (**C**). The best linear fit –continuous line– is close to the relation expected from mean-field theory, $\Delta_0/k_B T_c = 1.76$ –dashed line.

Supplementary Materials for

# A Family of Finite-Temperature Electronic Phase Transitions in Graphene Multilayers

Youngwoo Nam, Dong-Keun Ki, David Soler-Delgado, and Alberto F. Morpurgo

Correspondence to Alberto F. Morpurgo (alberto.morpurgo@unige.ch)

**This PDF file includes:**

## Materials and Methods

Device fabrication and identification of multilayer thickness and staking-order

Suspended graphene devices were fabricated on a sacrificial lift-off resist layer (LOR 10A, MicroChem) and were cleaned using a current annealing technique before performing systematic electrical measurements. Both two-terminal and four-terminal devices were employed in the course of our study (see the inset of Fig. S1 for schematics of the different configurations). The device fabrication process and the current annealing procedures have been extensively discussed multiple times in past publications from our group (*10, 18, 19, 49-52*) and we refer the reader to these papers for details. Here, we only briefly recall how the thickness of graphene multilayers is determined.

A key indicator of the multilayer thickness and stacking is provided by the first quantum Hall plateau appearing at low magnetic field, typically 0.1 T or less (i.e., at a magnetic field that is sufficiently low to prevent the effects of electron-electron interactions on Landau levels). For Bernal-staked $N$-multilayers, the first quantum-Hall plateau is expected to develop at a filling factor $\nu = 2N$, with a value of quantized transverse conductance $G_{xy} = 2Ne^2/h$ (*53*). We have indeed shown earlier that this is the case for all even multilayers up to $N = 8$ (*18, 19*). The current work is based on the analysis of transport through approximately 20 suspended multilayers in which the quantum Hall effect was consistent with this expected behavior. Devices whose Hall conductance quantization prevented a clear determination of the thickness were discarded from the analysis. In two-terminal devices, this can indeed happen due to the effect of the contact resistance. Only in one case, for a rather thick multilayer (as inferred from the optical contrast) we observed a good quantization of the Hall conductance incompatible with the one expected for Bernal-stacked multilayers (the Landau level fan diagram measured experimentally exhibited crossing of Landau levels that we never observed in any other suspended device). We attributed the phenomenon to a non-Bernal stacking of that specific multilayer, which was therefore not included in our analysis. The small percentage of non-Bernal stacked devices that we found in the experiments is consistent with the fact that, throughout our work, natural graphite was used to produce exfoliated multilayers.

The determination of thickness based on low-field quantum Hall effect did give consistent results when cross-checked with other methods. For instance, in some devices magneto-Raman measurements were employed to confirm the thickness and stacking of multilayers (*18, 40, 41*). Also the behavior of $n^*(T)$ discussed in the main text was extremely reproducible from sample-to-sample for a given thickness (so much so that at this stage of the work we can identify the thickness of a multilayer by measuring $n^*(T)$). Finally, during the last part of the work, we also started to exfoliate multilayers onto silicon/silicon oxide substrates, and to transfer them subsequently onto LOR (i.e., instead of exfoliating the multilayers directly on LOR). As we show in Fig. S1, a careful analysis of the optical contrast of multilayers on these substrates allows the discrimination of the multilayer thickness. These measurements do not exclude that the stacking of the multilayers is different from Bernal, but in view of the very small percentage of non-Bernal stacked multilayers that we found when exfoliating natural graphite, this possibility is not posing a significant hurdle in practice.

## Supplementary Text

Extracting $n^*(T)$ from the conductivity data: robustness of the determination of $T_c$ and $\Delta_0$

In the main text, we have shown that the density of carriers thermally excited from the valence to the conduction band at charge neutrality, $n_{th}(T)$, corresponds to the width of Dirac peak, $n^*(T)$, which is why the analysis of $n^*(T)$ allows the density of states of suspended graphene multilayers to be probed. This provides important information not accessible from the measurements of the temperature dependence of the resistance. In particular, the behavior of $n^*(T)$ allows us to prove the occurrence of a second-order phase transition in all Bernal-stacked graphene multilayers investigated, from bilayers to heptalayers.

In view of its relevance for our results, it is important to discuss in detail how $n^*(T)$ is determined experimentally. We will show that none of our key conclusions (namely the occurrence of a second-order phase transition, the value of the critical temperature, the value of the gap, and the fact that mean-field theory reproduces the behavior of $n^*(T)$ very satisfactorily) depend on details of how $n^*(T)$ is defined or obtained.

Our recent work on bilayer graphene (*52*) demonstrates that the experimentally measured conductivity near charge neutrality is perfectly reproduced quantitatively by the relation

$$\frac{\sigma(n,T)}{\sigma(n=0,T)} = \frac{1}{4gk_BT\ln(2)}\frac{(n_e+n_h)(n_e^2+n_h^2)}{n_en_h}. \qquad \text{(S1)}$$

Here $n_e(n,T) = gk_BT\ln\left[1+e^{E_F(n)/k_BT}\right]$ and $n_h(n,T) = gk_BT\ln\left[1+e^{-E_F(n)/k_BT}\right]$ are the density of thermally excited electrons and holes; $E_F(n)$ is the Fermi energy under accumulation of a density of charge carriers $n$ (positive $n$ and $E_F(n)$ correspond to electron accumulation; negative $n$ and $E_F(n)$ correspond to hole accumulation); $g = 2m^*/(\pi\hbar^2)$ is density of states (DOS) of the non-interacting bilayer graphene ($m^*$ is the effective mass in bilayer graphene). The density of electrons thermally activated from the valence to the conduction band at charge neutrality –which we indicate with $n_{th}(T)$– corresponds then to $n_{th}(T) = n_e(n = 0, T)$. It follows directly from Eq. (S1) that when $n = n_{th}(T)$, $\sigma(n = n_{th}(T),\ T) \cong 1.67\,\sigma(n = 0,\ T)$. Hence, experimentally, if we define the width, $n^*(T)$, of the conductivity-vs-density curve as the value of density $n$ for which the measured conductivity equals 1.67 times the conductivity at charge neutrality, we automatically have that $n^*(T) = n_{th}(T)$. The $n^*(T)$ data shown in Fig. 1G of the main text are obtained using this criterion.

In Fig. 1G of the main text, the continuous curve represents the calculated quantity

$$n_e(n=0,T) \equiv n_{th}(T) = \int_0^\infty g(\varepsilon,T)f_{FD}(\varepsilon,T)d\varepsilon, \qquad \text{(S2)}$$

where $f_{FD}(\varepsilon,T) = 1/(e^{\varepsilon/k_BT}+1)$, i.e. the Fermi-Dirac distribution function with $E_F = 0$.

For $T > T_c$, there are no free parameters. $n_{th}(T)$ equals to $gk_BT\ln(2)$ and depends only on $T$ and on the effective mass of bilayer graphene, which is equal to $m^*= 0.033\ m_e$ ($m_e$ is the free electron mass) (*52, 54-57*). It is seen that Eq. (S2) perfectly matches the data with no variable parameters.

For low $T$ –i.e., below the transition temperature– Eq. (S2) depends on the value of the gap since for $T < T_c$ the DOS is modified by interactions and becomes $g(\varepsilon, T) = \frac{2m^*}{\pi\hbar^2}\frac{\varepsilon}{\sqrt{\varepsilon^2-[\Delta(T)/2]^2}}$ . We assume a mean-field temperature dependence of the gap (approximated with a precision of 2% or better by the expression $\Delta(T) = \Delta_0 \tanh\left(1.74\sqrt{T_c/T - 1}\right)$ for all values of $T < T_c$) and we fit the data using $T_c$ and $\Delta_0 = \Delta(T = 0)$ as parameters. $T_c$ is readily identified by looking at the data and cannot be varied. Therefore $\Delta_0$ is the only parameter that can be varied freely. As it is apparent from Fig. 1G, the result of this analysis is fully consistent for four different devices, and theory reproduces the data perfectly.

We now show that none of the key results depend on the choice of the value 1.67, in the relation $\sigma(n = n^*,\ T) \cong 1.67\ \sigma(n = 0,\ T)$ that we used to extract the value of $n^*(T)$. To illustrate this point we show in Fig. S2A and Fig. S3A $n^*(T)$ for a bilayer and a trilayer device, respectively, extracted using different criteria, namely $\sigma(n = n^*,\ T) \cong a\ \sigma(n = 0,\ T)$ with different values of $a$ ranging from 1.1 to 2. As it is apparent from the figures, changing the value of $a$ results in a change of the slope of $n^*$-vs-$T$ at high temperature (i.e., for $T > T_c$). However, neither $T_c$ nor $\Delta_0$ are affected: precisely the same value of $T_c$ and $\Delta_0$ is found irrespective of the value of $a$ chosen (see Fig. S2B-C and Fig. S3B-C). To emphasize this conclusion, we plot the data extracted with the different criteria in terms of reduced variables, i.e., $n^*(T)/n^*(T_c)$ as a function of $T/T_c$. As shown in the insets of Fig. S2A and Fig. S3A when plotted in this way all data sets collapse on top of each other. This result clearly shows that the value of $T_c$ and $\Delta_0$ is independent of the value of $a$ used in the analysis.

We have checked that this insensitivity of our conclusions to the criterion used to extract $n^*(T)$ holds true for all multilayers investigated. This insensitivity is why our analysis is extremely robust. Note that it is important to check this, because for multilayers thicker than bilayers the criterion $\sigma(n = n^*,\ T) \cong 1.67\ \sigma(n = 0,\ T)$ has not been justified theoretically and small deviations are expected. Indeed, Eq. (S1) is exactly valid in bilayers, but is only approximately valid in thicker multilayers. The microscopic reason is that in multilayers thicker than bilayers more conduction and valence bands are present (see Appendix 1 and (*36-39*)). In such a situation, scattering between electrons (holes) in different conduction (valence) bands contributes to relax the velocity of charge carrier and to limit the conductivity. This effect is not present in bilayers since only one conduction and one valence band are present, and is therefore not included in Eq. (S1), which only considers the effect of electron-hole scattering. The discussion here nevertheless shows that these details play no role in the determination of $T_c$ and $\Delta_0$ .

Limits of the analysis of $n^*(T)$ at low temperature ($T \ll T_c$)

As temperature is lowered the value of $n^*(T)$ decreases. It may be expected that at sufficiently low temperature there can be difficulties in extracting a reliable value for $n^*(T)$ for two main reasons. The first is that if $T$ is decreased to low enough values, $n^*(T)$ should eventually become smaller than the inhomogeneity in carrier density originating from extrinsic disorder. At that point the value of $n^*(T)$ extracted from the data gives a measure of the carrier density induced by disorder, and not of the density of thermally excited carriers. The second reason is that the value of $n^*(T)$ is determined from a resistance measurement, under the implicit assumption that transport through the multilayer is uniform. However, in even multilayers at sufficiently low $T$ the square resistance becomes very large. Under these conditions it should be expected that the measured resistance is determined by the highest conducting percolating path between source and drain (very likely at the device edges) (*34, 35*). If so, transport does not occur uniformly through the entire system and the measured resistance does not allow to extract information about "bulk" properties of multilayers.

To assess if and when these issues pose problems we can simply plot the value of $n^*(T)$ determined from the experimental data and check when our theoretical expression matches the measured values. We have done this systematically for multilayers of all thicknesses (see Fig. S4 for a selection of examples) and found that the agreement between $n^*(T)$ and theory is excellent over a broad temperature range in all cases. Depending on the device, deviations start whenever $n^*(T) \sim 10^9$ cm$^{-2}$ or when the square resistance becomes comparable to $h/e^2$, precisely in line with our expectations. These deviations at low temperature do not affect any of our conclusions nor prevent a precise determination of the gap. Indeed, as $T$ is decreased below $T_c$, agreement with theory is seen in a broad range of temperatures in all cases. In this temperature range, the variation of $n^*(T)$ ranges from a factor of 20 to a factor of 500, depending on the multilayer thickness and on the specific device. This range of $n^*(T)$ value is comfortably broad to extract quantitatively the value of $\Delta_0$ in all devices.

$n^*(T)$ of monolayer graphene

The idea to analyze how the width of the resistance peak near charge neutrality depends on temperature to reveal the occurrence of phase transitions in multilayer graphene (i.e., what we discuss in the main text) is new. However, for suspended monolayer graphene a similar analysis has been done in the past to extract information about the density of states (*31*). The experiments on monolayers revealed the expected quadratic dependence of $n^*$ on $T$ ($n^* = \pi/6\,(k_B T/\hbar v_F)^2$), which originates from the fact that the monolayer DOS increases linearly with energy from the charge neutrality point. At a quantitative level, the analysis yielded a value of the Fermi velocity $v_F = 2 \times 10^6$ m/s larger than the value commonly taken for graphene, which was argued to agree with the expected renormalization of $v_F$ due to electron-electron interactions (*22, 46-48, 58, 59*).

Fig. S5 shows data from one of our monolayer devices in which we performed the same analysis. We find that $n^*(T)$ exhibits a smooth quadratic behavior with no evidence for a phase transition and a value of $v_F = 1.7 \times 10^6$ m/s enhanced as compared to the commonly expected one, in nearly perfect agreement with earlier work on monolayer (the red line in Fig. S5 is a fit to the data that we did to extract the value of $v_F$ using the relation $n_{th}(T) = \pi/6\ (k_B T/\hbar v_F)^2$). Note that the absolute value of $n^*(T)$ –for a same given temperature between 10 and 100 K– is two orders of magnitude smaller in monolayers than for thicker multilayers. The difference originates from the DOS associated to the linear graphene band, two orders of magnitude smaller (or more) than the DOS of thicker multilayers in the relevant energy range. This is also why the presence of the linear band can be safely neglected when analyzing the behavior of odd multilayers such as 3LG or thicker.

Earlier estimates of the gap

The analysis of the temperature dependence of $n^*(T)$ is directly sensitive to the DOS of Bernal-stacked multilayers. It is this fact –together with the excellent reproducibility of the data and the very systematic behavior observed upon varying the multilayer thickness– that allows us to extract a reliable value for the gap $\Delta_0$ that opens at charge neutrality. In contrast, earlier attempts by different groups (including ours) to determine the magnitude of the gap in suspended graphene multilayers relied on either bias dependent conductance measurements or on the value of the activation energy extracted from the resistance (*15-19, 23, 60*). As we now discuss, these estimates are incorrect and the conclusions obtained from their analysis erroneous.

To discuss the case of bias-dependent differential conductance measurements we first refer to earlier experiments from our group on even Bernal-stacked multilayers (up to 8LG) (*18, 19*). Fig. S6A-D show color plots of the low-temperature differential conductance measured as a function of gate and bias voltage on 2LG, 4LG, 6LG, and 8LG devices, in which a pronounced suppression is seen in all cases. Fig. S6E shows cuts of these color plots at charge neutrality: upon increasing the bias a sharp increase in conductance is observed at a sharply defined threshold. The corresponding energy may be taken (and has been taken in different cases in the past) as a measure of the gap. However, Fig. S6F shows that the values obtained in this way (red empty circles) are in general much smaller than the values of $\Delta_0$ extracted from the analysis of $n^*(T)$ (empty black squares). This comparison directly shows that the energy scale probed with bias-dependent measurements is not the gap (most likely it is a manifestation of Coulomb blockade affecting the percolating path responsible for hopping-mediated transport that dominates when devices are highly resistive).

Attempts to use this type of measurements to estimate a critical temperature are similarly unreliable. This can be concluded, for instance, by looking at earlier work on suspended ABC-stacked trilayers, in which a very strongly insulating state is also seen (*60*). In this case, bias-dependent measurements and their temperature evolution were used to argue that the value of the gap is $\Delta_0 \sim 43$ meV and that $T_c = 34$ K. This result is not at all internally consistent since $\Delta_0 \sim 15\ k_B T_c$ whereas it is expected that $\Delta_0 \sim\ k_B T_c$.

Finally, attempts to estimate the gap from the temperature dependence of the conductance at charge neutrality are also problematic. For the same devices whose data are shown in Fig. S6A-D, the activation energy is plotted in Fig. S6F as blue empty triangles. It is seen that the activation energy is nearly thickness independent and always close to $\Delta_0 \sim 1.5$ meV (*19*) (comparable to the value obtained from the bias dependent measurements). Most likely this is again because the activation energy is probing the contribution to transport of the dominating percolating path (*34, 35*), and not of the "bulk" of the multilayer. An estimation based on the activation energy is also difficult, because the value extracted can depend sensitively on the temperature range used for the analysis of the data.

In conclusion, the analysis of $n^*(T)$ gives a reproducible way to extract the gap and the critical temperature that is internally consistent, something that could never be done reliably until now.

<u>Finite-temperature versus quantum phase transition</u>

The data in the main text show clearly a sharp change in the temperature dependence of $n^*(T)$, from linear to sublinear, which allows us to identify the value of the critical temperature $T_c$. This implies the occurrence of a finite temperature phase transition. However, one may wonder whether the data are sufficiently clear to discriminate this conclusion from a scenario in which the insulating state is due to a $T = 0$ quantum phase transition. In such a case, at $n = 0$ a gap is present for all temperatures and does not close at $T_c$. To assess this possibility, we calculate the expected $T$-dependence of $n_{th}(T)$ and compare it to the data. This is done in Fig. S7 for 3LG (other thicknesses would lead to the same result). Panel A shows the analysis discussed in the main text based on a finite temperature second-order phase transition with $\Delta(T)$ vanishing at $T_c$. Panel B shows the case of a $T$-independent gap expected from a quantum phase transition. It is apparent that this latter scenario cannot reproduce the data satisfactorily.

<u>Crossover to the behavior of bulk graphite</u>

The occurrence of a phase transition in thick Bernal-stacked multilayers with $T_c$ increasing upon increasing thickness is particularly striking because one would expect to recover the semi-metallic behavior of graphite in sufficiently thick multilayers. Our experiments provide (for the first time) a clear indication as to the mechanism responsible for the crossover to the behavior of bulk graphite, and of the thickness for which this crossover occurs. The idea is illustrated in Fig, S8.

We have shown in the main text that upon increasing thickness, the magnitude of the gap that opens in the quadratic bands increases systematically, and exhibits a linear dependence on the number of layers. In Fig. S8 this linear dependence is extrapolated for a generic $N$-multilayer graphene (black line in Fig. S8). We see that when at a thickness of approximately 30 monolayers, $\Delta_0$ becomes as large as the parameter $\gamma_\perp$, the (smallest) effective hopping amplitude in the direction perpendicular to the plane. $\gamma_\perp$ determines how far away in energy the "high-energy" bands are from the charge neutrality (see the band diagram for Bernal-stacked multilayers plotted in Fig. S10 and discussed in Appendix 1). These bands are normally far enough in energy to be neglected, so that the behavior of thin multilayers can be described in terms of an effective low-energy model (*36-39*). Our phenomenological description of the phase transition in Bernal-stacked multilayers in terms of a staggered potential is valid in this regime (*18, 19*). Fig. S8 implies that an extrapolation of the behavior observed up to 7LG will certainly break down at a thickness of the order of 30 layers because starting from that thickness the “high-energy bands” cannot be neglected.

## Mean-field $T_c$ formula

It will be essential to develop a microscopic theory of the broken symmetry state that we observe to understand all details of its evolution upon increasing multilayer thickness. In the absence of a microscopic theory, however, we try a first analysis of the very systematic experimental observations using the simplest possible ideas that are motivated by the observed phenomenology. In particular, as we have discussed, all data appear to be systematically consistent with the behavior expected from mean-field theory. Quite generically, within such a mean-field approach, the expression for the critical temperature reads

$$k_B T_c = \frac{E_{cut-off}}{2\sinh[1/(gV)]} \qquad \text{(S3)}$$

where $E_{\text{cut-off}}$ and $V$ represent the energy cut-off and strength of the interaction, respectively ($k_B$ is the Boltzmann constant). $g$ is the density of states that is energy independent, which is known from the known band structure of multilayer graphene (see Appendix 1; $g$ and is determined by the sum of the effective masses of the different quadratic bands). For bilayers, Eq. (S3) –virtually identical to the equation of BCS theory of superconductivity (*61*)– can indeed be obtained from a microscopic theory.

If we assume that the interaction strength $V$ is the same for all multilayers investigated (which may be the case if the range of the interaction is much larger than the thickness of the thickest layer studied), we can check whether Eq. (S3) reproduces the measured $T_c$ data. Fig. S9 shows that it does. The red line represents Eq. (S3) with $E_{\text{cut-off}} \sim 120$ K and $V \sim 1.8 \times 10^{-10}$ cm$^2$K and shows that all available data points are compatible with this expression for $T_c$. As we made clear earlier, it remains to be verified theoretically if an approach of this type is physically meaningful and whether the assumptions made (e.g., whether considering that $V$ is thickness independent) are correct. Nevertheless, we find it worth discussing this comparison explicitly because it makes an important point clear. It illustrates that having access to an entire family of related phase transitions, exhibiting a systematic evolution with thickness, allows new tests that are not possible in other systems investigated in the past. These tests, and the underlying systematics of the experimental observations, both pose tight constraints on any possible theory and provide good indications for the relevant microscopic aspects that a theory has to capture.

## Appendix 1. Band structure of graphene multilayers

In this section, we recall the key aspects of the band structure of Bernal-stacked multilayers obtained from a minimal tight-binding model (i.e., considering only the nearest neighbor intra- and interlayer layer hopping parameters $\gamma_0$ and $\gamma_1$), which form the basis of the theoretical analysis of their electronic properties. We discuss features and quantities that are particularly relevant for the work discussed in this paper and we refer to the literature (*36-39*) for details.

Within the minimal tight-binding model the complete band structure of a generic Bernal-stacked *N*-multilayers can be decomposed into linear and quadratic bands. Each linear band has the same character of the Dirac band in monolayer graphene and each quadratic band is analogous to the quadratic bands present in bilayer graphene (with a different mass). In the absence of a gap the system is perfectly electron-hole symmetric (which is indeed also the case for mono and bilayer graphene) and all the bands touch at $E = 0$ (i.e., at the charge neutrality point).

More specifically, at low-energy, any odd $N = 2M+1$ Bernal-stacked multilayer possesses one linear conduction band and $M$ quadratic bands with different effective masses $m_n^*$ ($n$ = 1, 2, …, $M$; because of electron hole symmetry the same is true for the valence bands). For even $N = 2M$, only $M$ quadratic bands are present (see Fig. S10). The full set of bands from bilayers to 7LG is shown in Fig. S10. The effective mass of individual $n$-th quadratic band in $N$-multilayer is given by $m_n^* = \gamma_\perp^{N,n}/2v_F^2$, where $v_F$ is the Fermi velocity of monolayer. Using the known value $\gamma_1$= 0.39 eV, we can calculate the sum of effective mass of all quadratic bands in $N$-multilayer, $m^* = \sum_{n=1}^{M} m_n^*$ which is the quantity that determines the total DOS of each multilayer entering the expression for $n_{th}(T)$ discussed in the main text. $m^* = \sum_{n=1}^{M} m_n^*$ increases from $0.033m_e$ for $N = 2$ to $0.14m_e$ for $N = 7$.

The diagrams include also the "high energy bands" that are split away from $E = 0$ by an amount equal to the effective interlayer hopping energy $\gamma_\perp^{N,n} = 2\gamma_1 \cos\left(\frac{n\pi}{N+1}\right)$. Specifically, all the low-energy quadratic bands are degenerate at zero-energy; to each low-energy quadratic band in an $N$-multilayer corresponds a "high-energy band" shifted by $\pm\gamma_\perp^{N,n}$ from zero-energy (the sign + and – corresponds to the "high-energy" conduction and valence bands respectively). Note that both for $N = 2M$ and for $N = 2M + 1$ (i.e., for even and odd multilayers), the lowest energy of these "high-energy bands" is obtained for $n = M$. By direct substitution $\gamma_\perp^{N,M} = 2\gamma_1 \cos\left(\frac{M\pi}{2M+1}\right)$ or $\gamma_\perp^{N,M} = 2\gamma_1 \cos\left(\frac{M\pi}{2M+2}\right)$ in the two cases, which tend to zero as $M$ becomes very large (see Fig. S8 and S10). In other words, for sufficiently thick multilayers these bands stop being at "high-energy": they affect the low-energy behavior of the system and cannot be neglected.

As a final remark, note that low-energy bands present in all graphene multilayers thicker than monolayers are quadratic only approximately: mathematically the dispersion relation is hyperbolic and a quadratic dispersion is only a (very good) approximation that holds true at sufficiently low energy. At the highest energy explored in our work (corresponding to the larger temperature reached in the experiments, $T < 150$ K), the quadratic approximation is excellent. Nevertheless, given the relevance of this point for our analysis, we have calculated the density of thermally excited electrons in the conduction band of charge neutral bilayers with the exact hyperbolic dispersion and compared it to the approximate quadratic one. The result is shown in Fig. S11 for bilayer graphene (for thicker multilayers the situation is entirely analogous). The deviation between the exact calculation with hyperbolic dispersion and the approximate one with parabolic bands is at most a few percent at the largest temperature. In practice therefore the effect is entirely irrelevant for our experiments except possibly in the thickest multilayers at the largest temperature. In that case this deviation may account for the small deviation from linearity seen in the data.

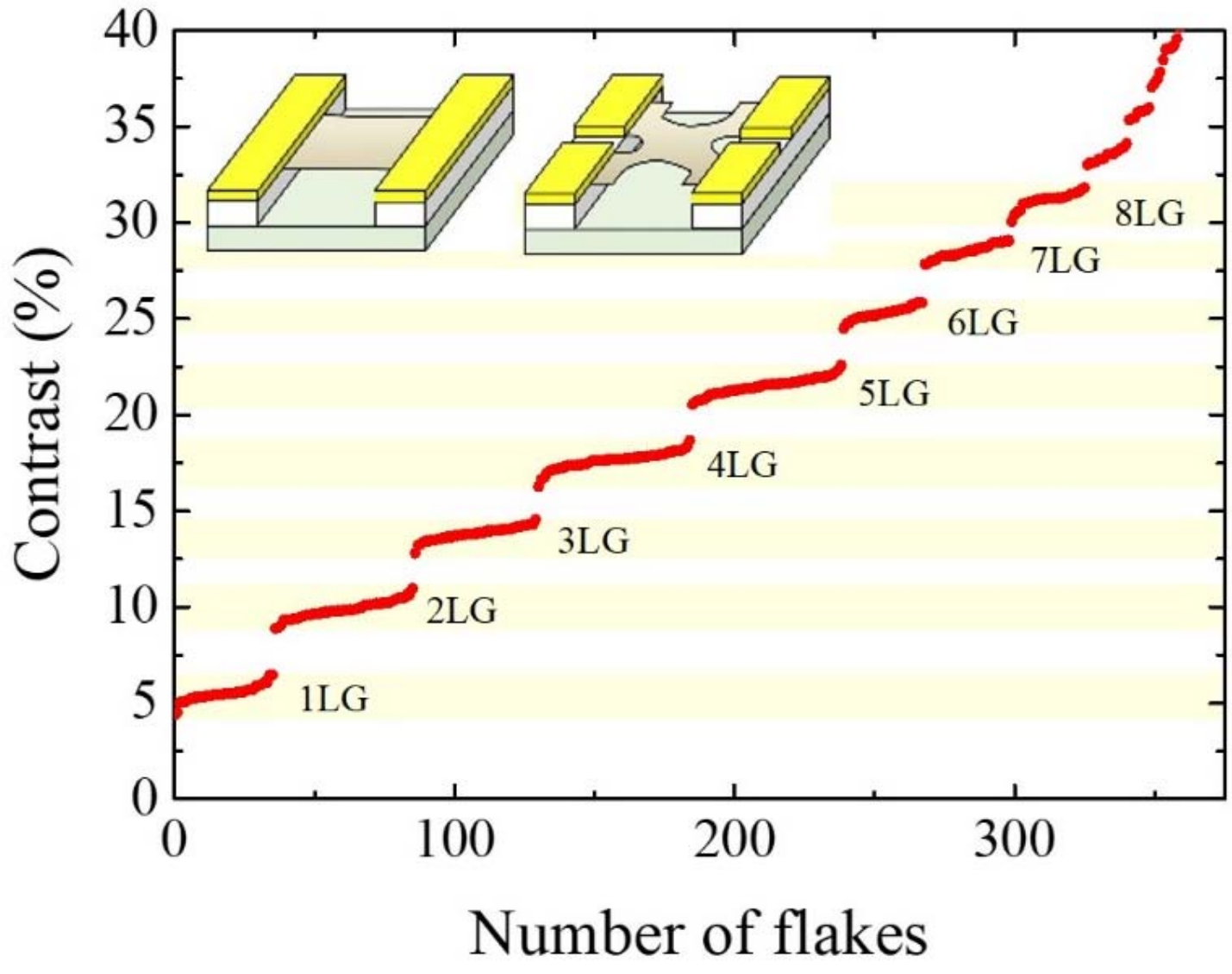

**Fig. S1.**

**Optical contrast of graphene multilayers on silicon/silicon oxide substrates.** We have measured the optical contrast of nearly 400 graphene multilayers exfoliated onto silicon substrates covered by 285 nm of thermally grown $SiO_2$. Measurements were done using a same microscope under identical conditions (light intensity, magnification, etc.) with the goal to discriminate between the layers of different thickness (*N*). Whereas most earlier attempt had not succeeded in discriminating layers thicker than 3-4LG, here we show that careful measurements allow the thickness of multilayers to be discriminated reliably up to much larger values. This is demonstrated by the quantization of the contrast that we observe at least up to 10LG. The method can likely work even for thicker layers, but our statistics for $N > 10$ is insufficient. The inset shows two-terminal and multi-terminal measurement configurations employed throughout this work.

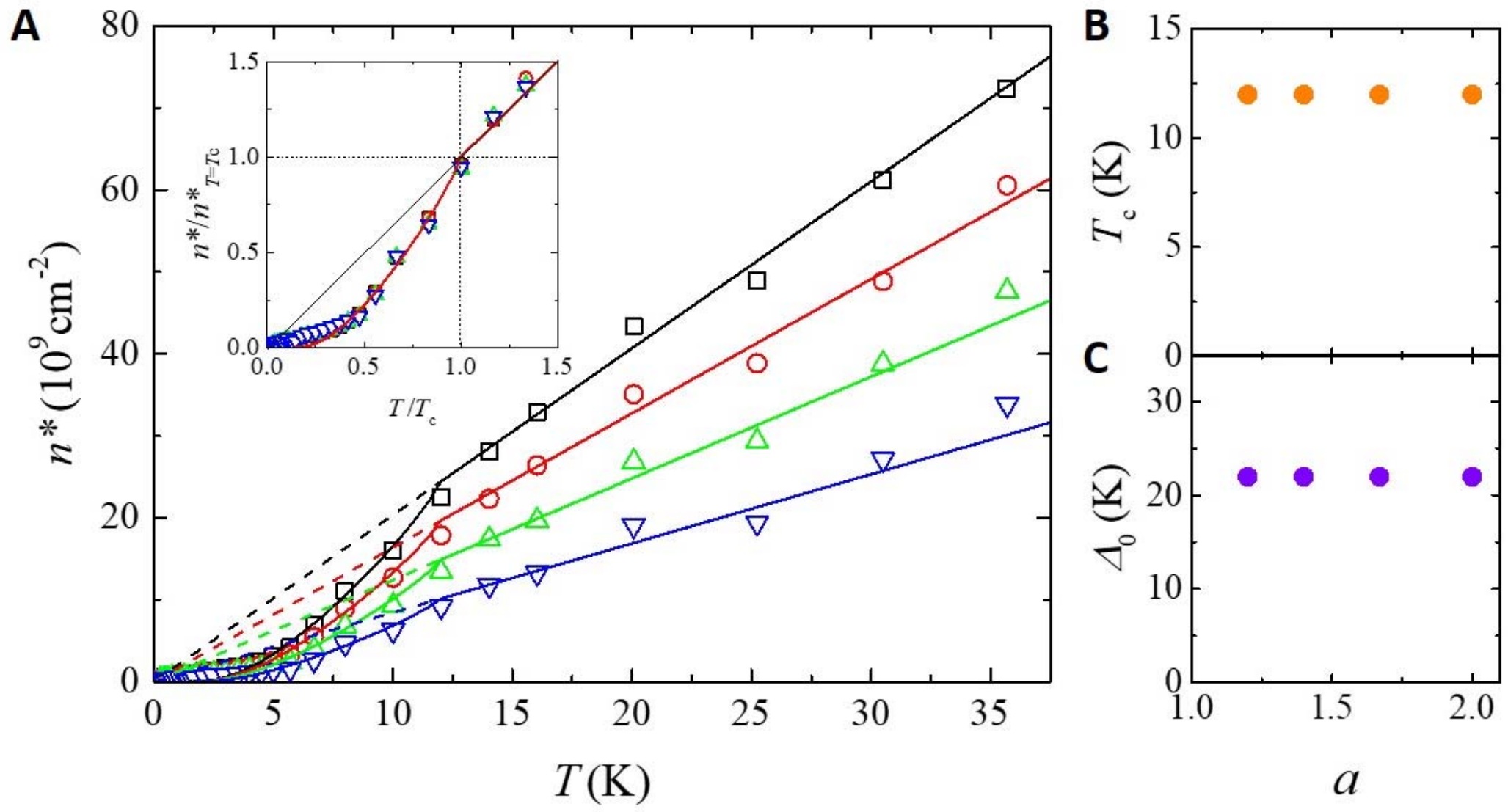

**Fig. S2.**

**Determination of the critical temperature $T_c$ and the gap $\Delta_0$ in bilayer graphene.** The determination of $T_c$ and $\Delta_0$ relies on the analysis of the temperature dependence of $n^*(T)$. As mentioned in the main text, the values of these quantities are insensitive to the criterion used to define $n^*(T)$. Here we illustrate this point in detail. (**A**) The different symbols represent $n^*(T)$ obtained from the density dependence of the conductivity using the criterion $\sigma(n = n^*,\ T) \cong a\,\sigma(n = 0,\ T)$ to define $n^*(T)$, with $a = 1.2$ (down triangles), 1.4 (up triangles), 1.67 (circles), and 2 (squares). The continuous lines of the corresponding color are fits based on a mean-field temperature dependence of the gap, as discussed in the main text and in the supplementary materials. Upon changing the value of $a$ the extracted values of $T_c$ and $\Delta_0$ remain unchanged as summarized by the plots in (**B**) and (**C**). The only quantity that changes is the slope of $n^*(T)$ for $T > T_c$. The insensitivity of $T_c$ and $\Delta_0$ to the chosen value of $a$ is emphasized when plotting the data in terms of reduced variables, i.e., when plotting $n^*(T)/n^*(T_c)$ versus $T/T_c$. As shown in the inset: all curves collapse on top of each other showing directly the insensitivity to the value of $a$.

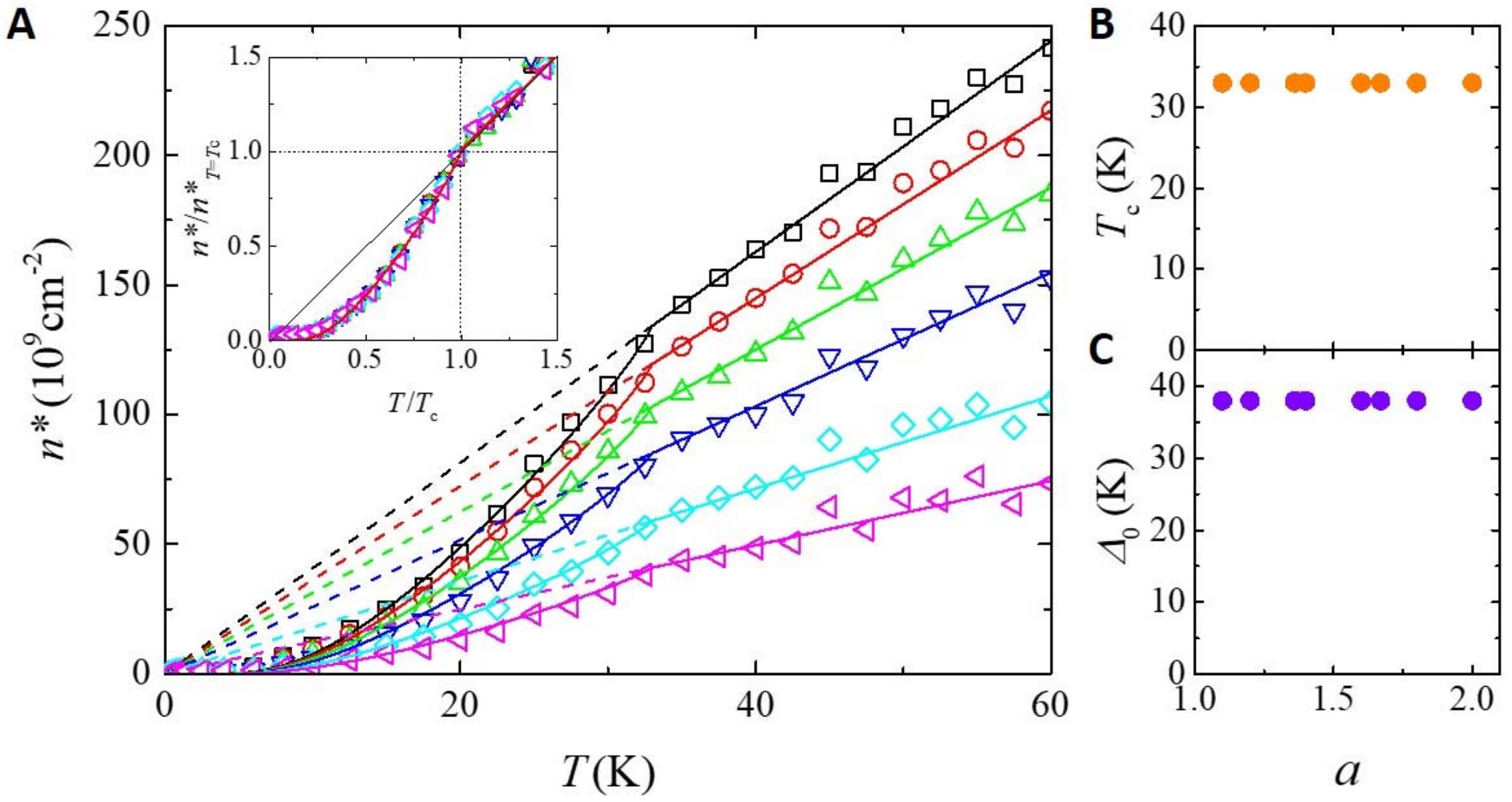

**Fig. S3.**

**Determination of the critical temperature $T_c$ and of the gap $\Delta_0$ in trilayer graphene.** The insensitivity of $T_c$ and $\Delta_0$ to the definition of $n^*(T)$ holds true not only for bilayer graphene, but also for all multilayer thicknesses. Here we confirm this result with data from graphene trilayers. (**A**) The different symbols represent $n^*(T)$ obtained from the density dependence of the conductivity using the criterion $\sigma(n = n^*,\ T) \cong a\,\sigma(n = 0,\ T)$ to define $n^*(T)$, with $a$ = 1.1 (purple triangles), 1.2 (diamonds), 1.4 (blue triangles), 1.6 (green triangles), 1.8 (circles), and 2 (squares). The continuous lines of the corresponding color are fits based on a mean-field temperature dependence of the gap, as discussed in the main text and in the supplementary materials. Just as for bilayers, upon changing the value of $a$ the extracted values of $T_c$ and $\Delta_0$ remain unchanged as summarized by the plots in (**B**) and (**C**). The only quantity that changes is the slope of $n^*(T)$ for $T > T_c$. The insensitivity of $T_c$ and $\Delta_0$ to the chosen value of $a$ is emphasized when plotting the data in terms of reduced variables, i.e., when plotting $n^*(T)/n^*(T_c)$ versus $T/T_c$. As shown in the inset: all curves collapse on top of each other showing directly the insensitivity to the value of $a$.

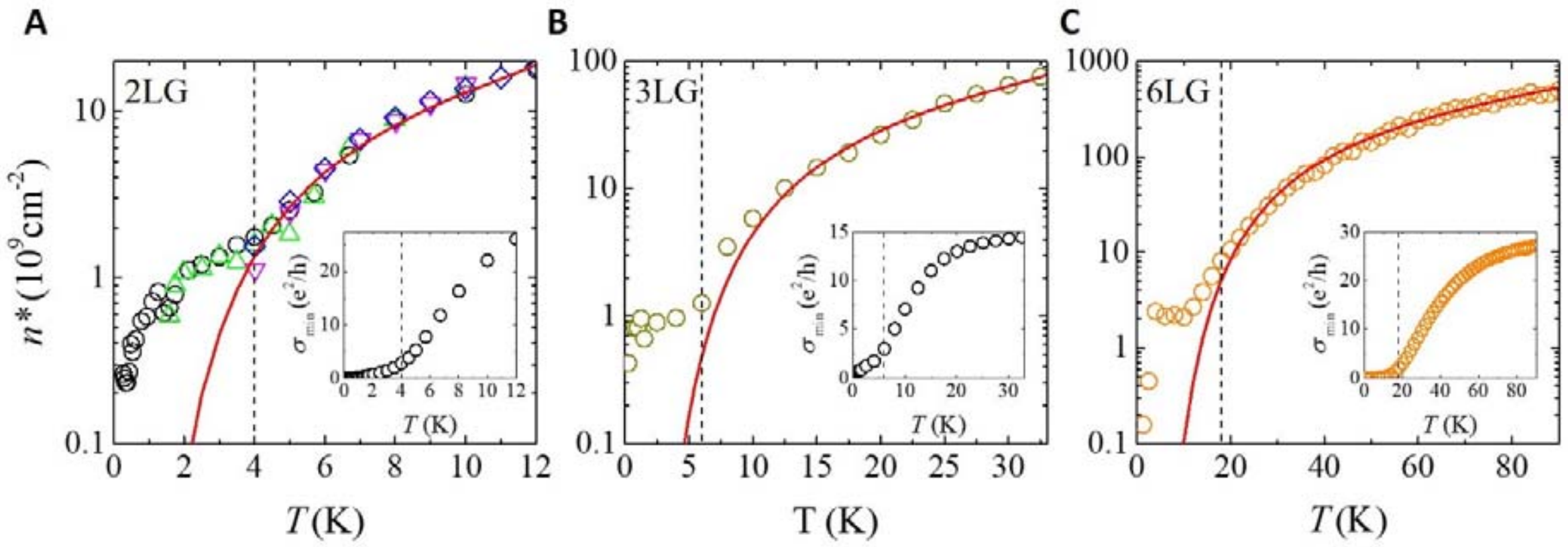

**Fig. S4.**

**Low-temperature behavior of $n^*(T)$ ($T \ll T_c$).** As temperature is reduced to values much smaller than $T_c$, the thermally excited density of carriers is expected to become very small due to the exponential suppression caused by the opening of the gap. As discussed in the supplementary materials, under these conditions the quantity $n^*(T)$ extracted from the gate dependent conductivity will stop corresponding to the density of charge carriers thermally excited from the valence to the conduction band. Here we analyze our experimental data to determine when this happens. We also show that the temperature interval over which $n^*(T)$ does correspond to the density of thermally excited charge carriers is sufficiently broad to allow the reliable extraction of $\Delta_0$. The symbols in panels (**A**)-(**C**) show the value of $n^*(T)$ determined experimentally for 2LG, 3LG, and 6LG respectively (the insets show the corresponding temperature dependence of the conductivity measured at charge neutrality). The red continuous line is calculated assuming that the gap has a mean-field temperature dependence. The vertical dashed lines indicate the value of temperature below which $n^*$ starts deviating from the calculated expected behavior. This analysis shows that deviations occurs when $n^*(T)$ is of the order of $10^9$ cm$^{-2}$ or when the conductivity becomes comparable to $e^2/h$. This behavior is caused by the presence of a small amount of extrinsic impurities or by the occurrence of inhomogeneous transport at the onset of the strongly insulating state of even multilayers. This is precisely what has been explained in the supplementary materials.

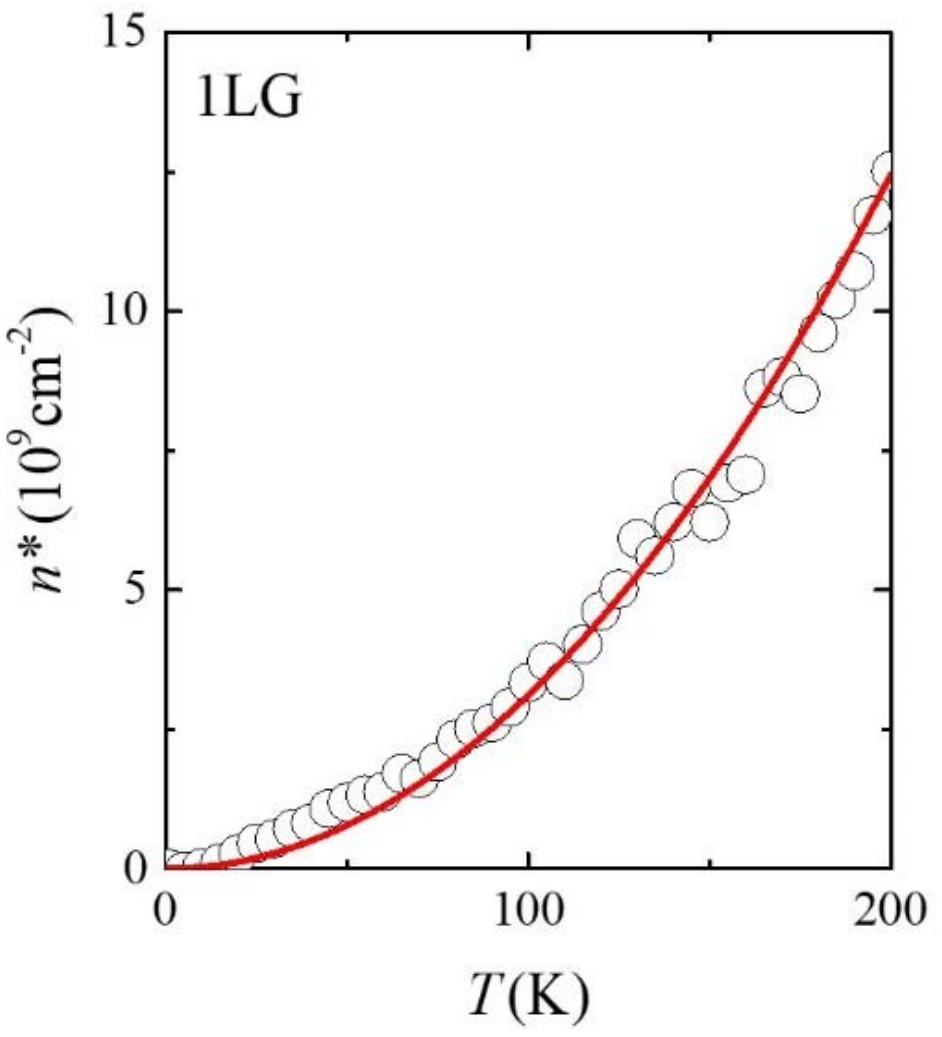

**Fig. S5.**

**Analysis of $n^*(T)$ for monolayer graphene**. The empty circles represent $n^*(T)$ measured on a graphene monolayer device, exhibiting a quadratic dependence on temperature $T$, with no evidence for a phase transition. The quadratic behavior originates from the monolayer DOS that is linear in energy and vanishes at $E = 0$ (i.e., at charge neutrality). The red continuous line is a fit to the data with the expression for the dependence expected for non-interacting Dirac electrons, $n_{th}(T) = \pi/6\,(k_B T/\hbar v_F)^2$. The data are reproduced using as value for the Fermi velocity $v_F = 1.7 \times 10^6$ m/s, larger than the value expected for graphene. This enhancement of $v_F$ –observed in similar earlier work– is a manifestation of renormalization due to interactions. Note that for, a same given temperature between 10 and 100 K, the absolute value of $n^*(T)$ is much smaller in monolayers than in thicker multilayers. That is why the total contribution to the DOS of the linear Dirac band present in odd Bernal-stacked multilayers is negligible.

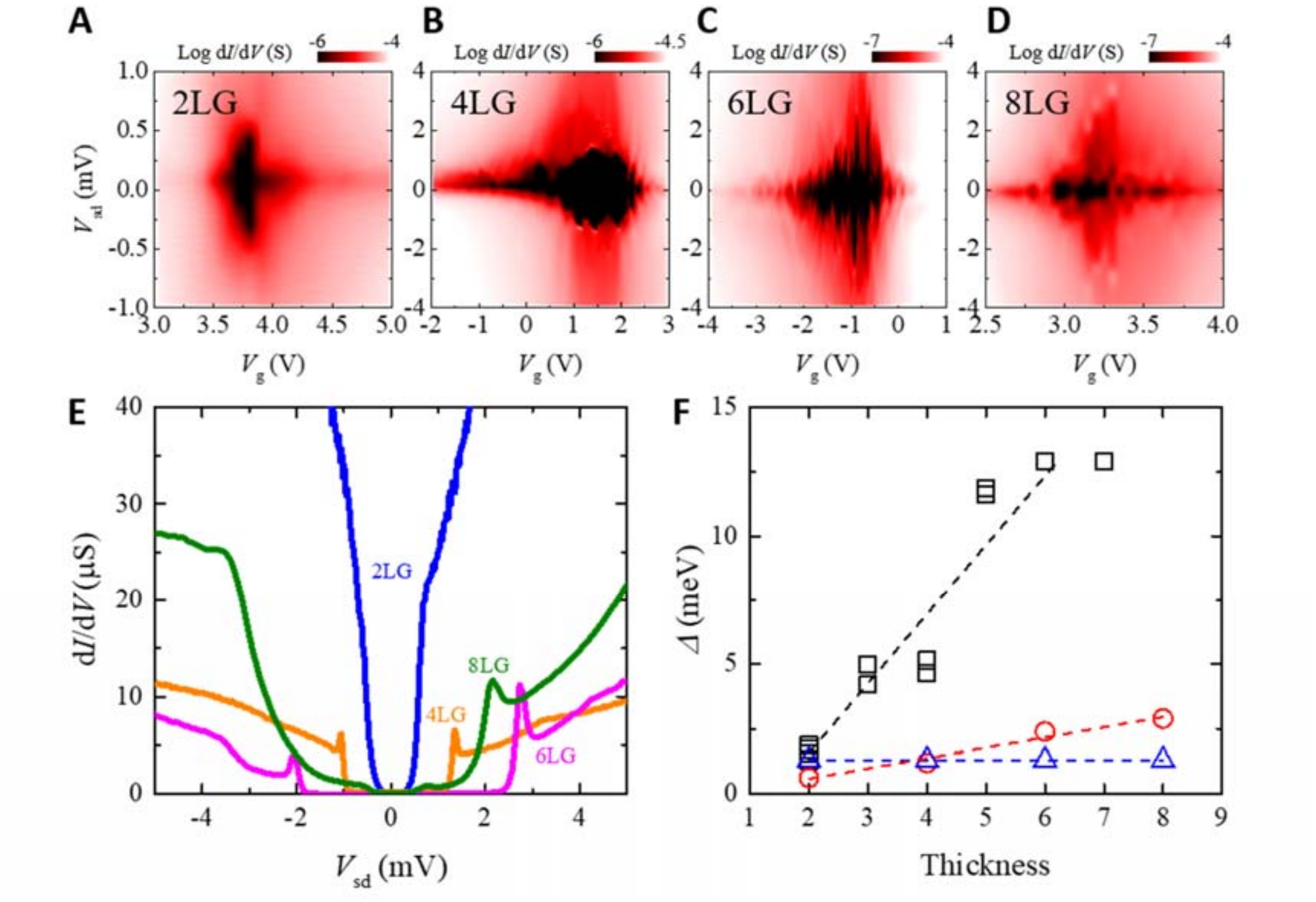

**Fig. S6.**

**Earlier attempts to estimate the interaction-induced gap in graphene multilayers.** Color plots showing the differential conductance d*I*/d*V* measured as a function of bias voltage ($V_{sd}$) and gate voltage ($V_g$) based on 2LG (**A**), 4LG (**B**), 6LG (**C**), 8LG (**D**) ($T$ = 250 mK). In all devices a pronounced suppression is visible around charge neutrality (black area). (**E**) Line cuts of the color plots at $V_g$ = 3.8, 1.4, -0.92, 3.15 V for devices based on 2LG, 4LG, 6LG, 8LG, respectively. Upon increasing the bias voltage, a sharp increase in conductance is observed: in the past, the thresholds for this increase have been used to estimate the gap. (**F**) Comparison of the gap in multilayer devices estimated from *i*) the analysis of $n^*(T)$ discussed in the main text (empty squares), *ii*) the bias-dependent conductance measurement (empty circles) shown in (E), and *iii*) the activation energy extracted from temperature dependence of the minimum conductance measured on the same devices (empty triangles; see Ref. (*19*)). The values of gap obtained from the bias dependence and the activation energy exhibit large deviation from (and are in general much smaller than) the values of the gap extracted from the analysis of $n^*(T)$ discussed in the main text.

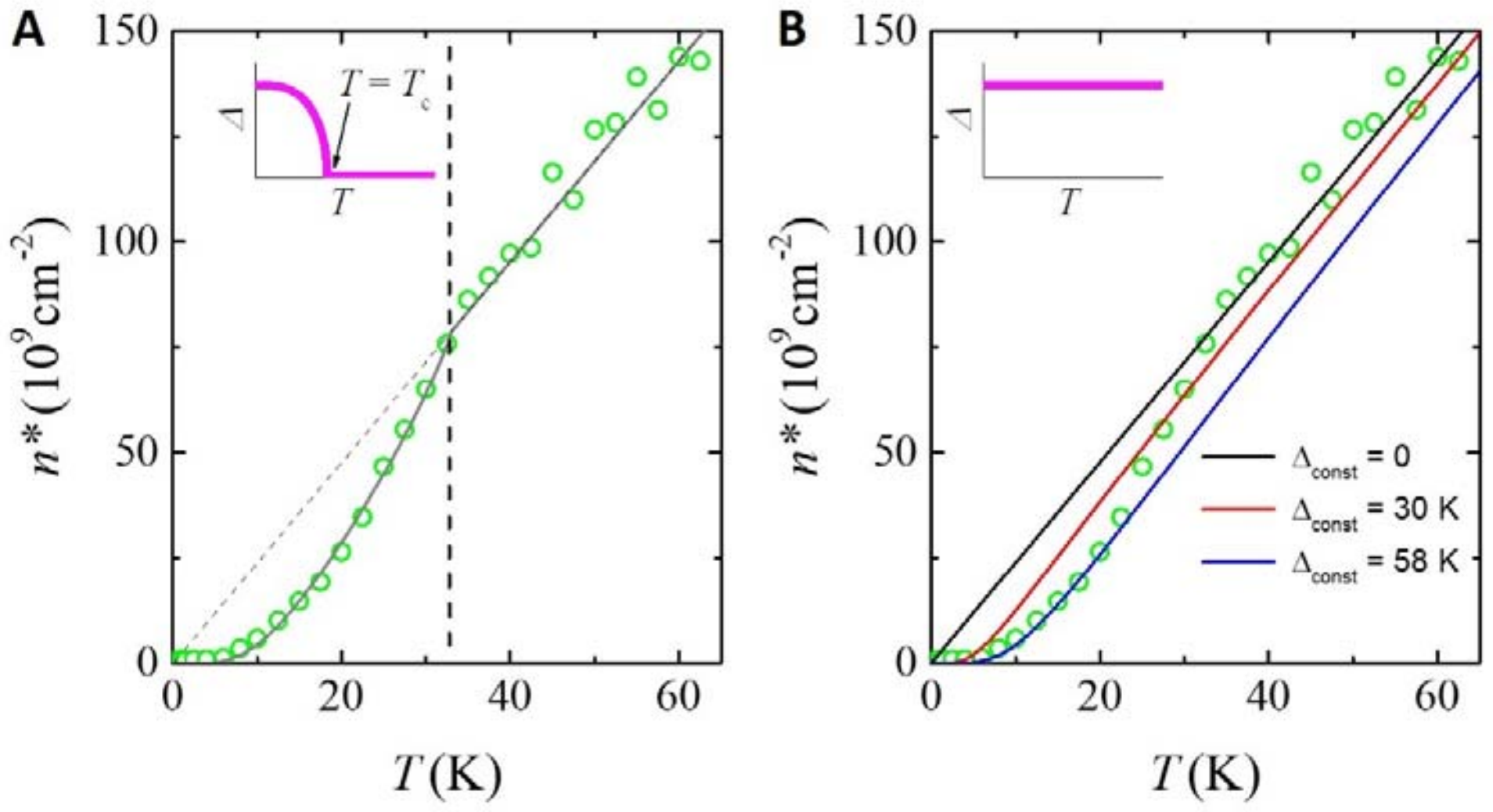

**Fig. S7.**

**Excluding the possibility of a quantum phase transition in Bernal-stacked multilayers.** Panel (**A**) shows experimental data (empty circles) for $n^*(T)$ of a trilayer graphene device, as discussed in the main text. The continuous grey line represents $n_{th}(T)$ calculated under the assumption that a finite-temperature second-order phase transition occurs, with a gap $\Delta$ exhibiting a mean-field temperature dependence and vanishing at $T_c$ (as shown in the inset). In panel (**B**) the same data are compared to the density of charge carrier thermally excited from the valence to the conduction band, under the assumption that the gap is constant (i.e., temperature independent, see the inset), as expected if the insulating state results from a $T = 0$ quantum phase transition. Lines of the different colors correspond to the case $\Delta = 0$ K (black line), 30 K (red line) and 58 K (blue line)). This comparison clearly shows that a constant, temperature independent gap cannot reproduce the data satisfactorily throughout the entire temperature range.

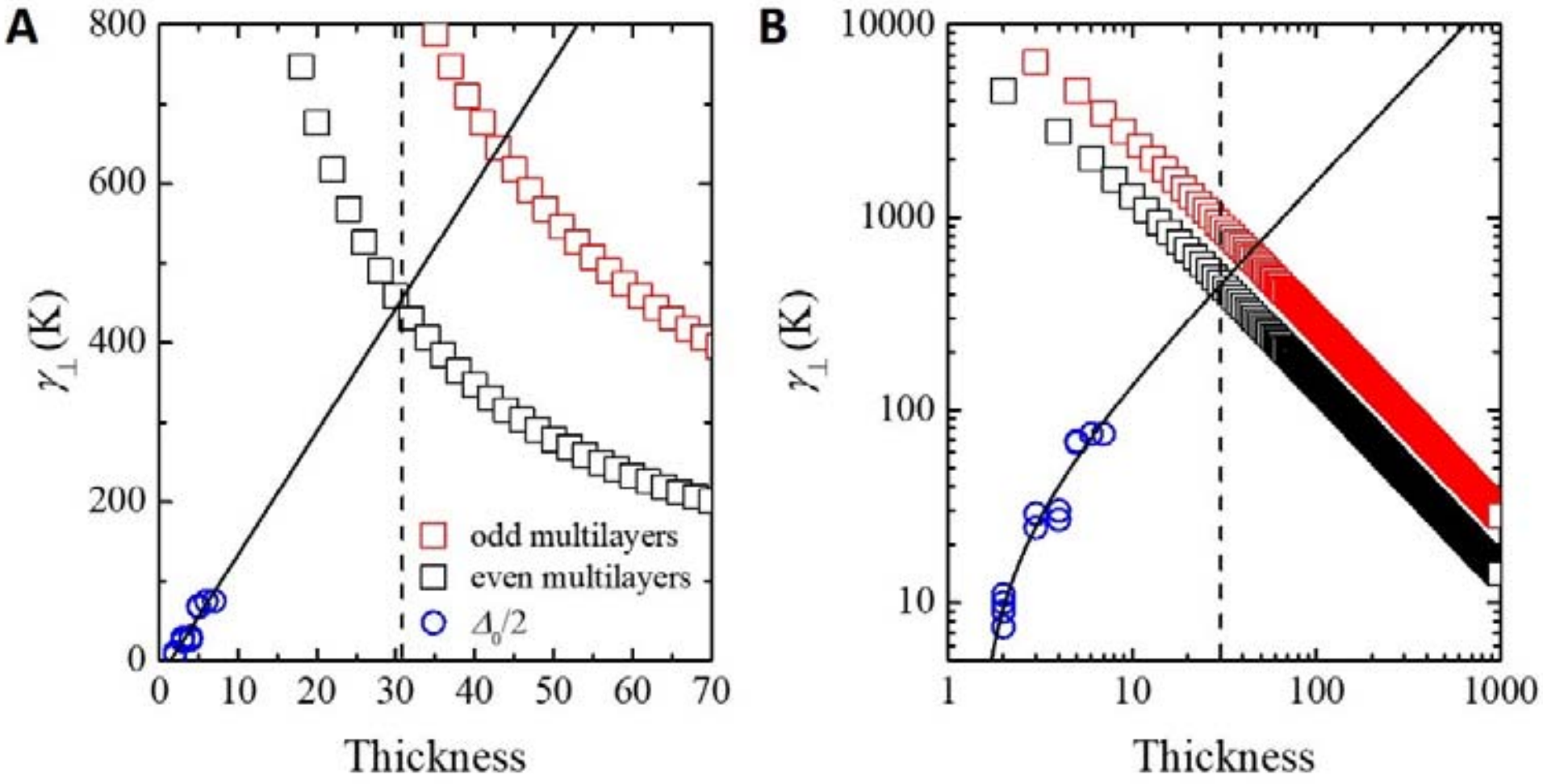

**Fig. S8.**

**Crossover to the behavior of bulk graphite.** (**A**) The black and red squares represent the theoretical value (see Appendix 1) of the smallest effective hopping amplitude in the direction perpendicular to the layer plane, $\gamma_\perp$, as a function of multilayer thickness (black and red symbols correspond to the case of even and odd multilayers, respectively). The blue empty circles correspond to $\Delta_0/2$, half the value of the experimentally determined gap $\Delta_0$ that opens at charge neutrality in all quadratic bands. The continuous line is a linear fit. Panel (**B**) shows the same plots in double-logarithmic scale. The parameter $\gamma_\perp$ determines how far away in energy from the charge neutrality are the "high-energy" bands (see Appendix 1) present in Bernal-stacked multilayers. Upon increasing thickness $\gamma_\perp$ decreases whereas the value of $\Delta_0$ increases. An extrapolation of this trend shows that the quantities become comparable for multilayer that are approximately 30 monolayer thick (as indicated by the vertical dashed line). For thicker layers, the "high energy" bands are sufficiently low in energy to "enter" the gap that opens between the low-energy quadratic bands. At these thickness values, the effect of the "high-energy" bands on the low-energy properties of multilayers cannot be neglected, i.e. the "low-energy" approximation explained in the main text certainly breaks down for thicknesses of the order of 30 layers. We expect that the lowering of the "high energy bands" is the main mechanism governing the crossover from the behavior of Bernal-stacked multilayers, which exhibit the opening of gap at charge neutrality, to that of graphite, which is a semi-metal.

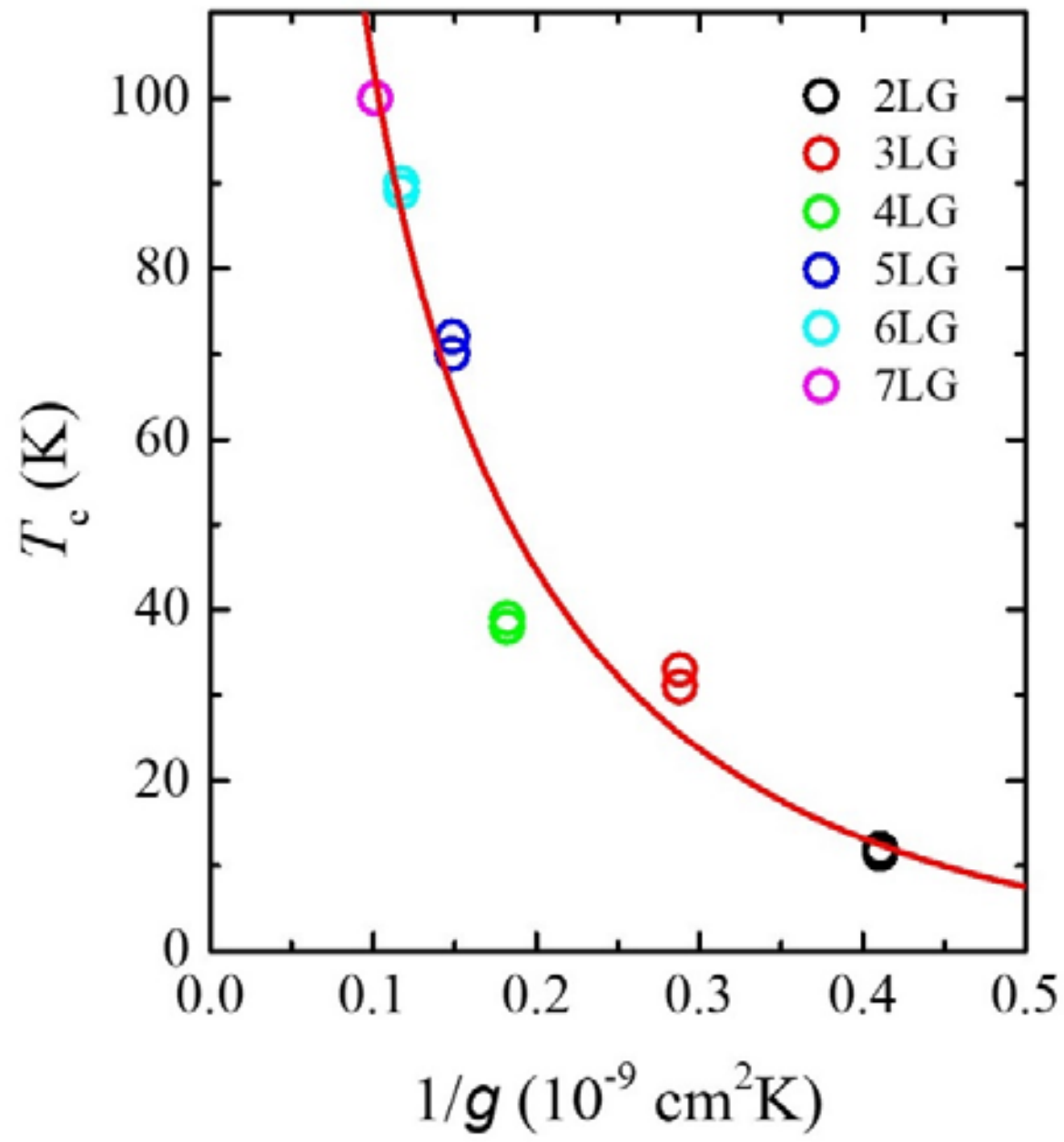

**Fig. S9.**

**Analysis of critical temperatures of multilayers within a mean-field theory.** Experimentally measured values of $T_c$ (empty circles) for multilayer plotted as a function of the inverse density of states at charge neutrality expected from theory (determined by the sum of the masses of all quadratic bands, see Appendix 1). The red line is obtained by using phenomenologically the theoretical mean-field expression for $T_c$, i.e., $k_B T_c = \frac{E_{cut-off}}{2\sinh[1/(gV)]}$ under the assumption that the interaction strength $V$ is the same for all multilayers in the range of thickness investigated. By treating $V$ and $E_{cut\text{-}off}$ as fitting parameters, we find that the expression reproduces well all data points with $E_{\text{cut-off}} \sim 120$ K and $V \sim 1.8 \times 10^{-10}$ cm$^2$K. Although a well-defined microscopic theory of interacting electrons in Bernal-stacked multilayers should be developed to justify this approach, this analysis is interesting because it shows how the evolution of measured quantities with thickness gives useful information to identify the correct theoretical model.

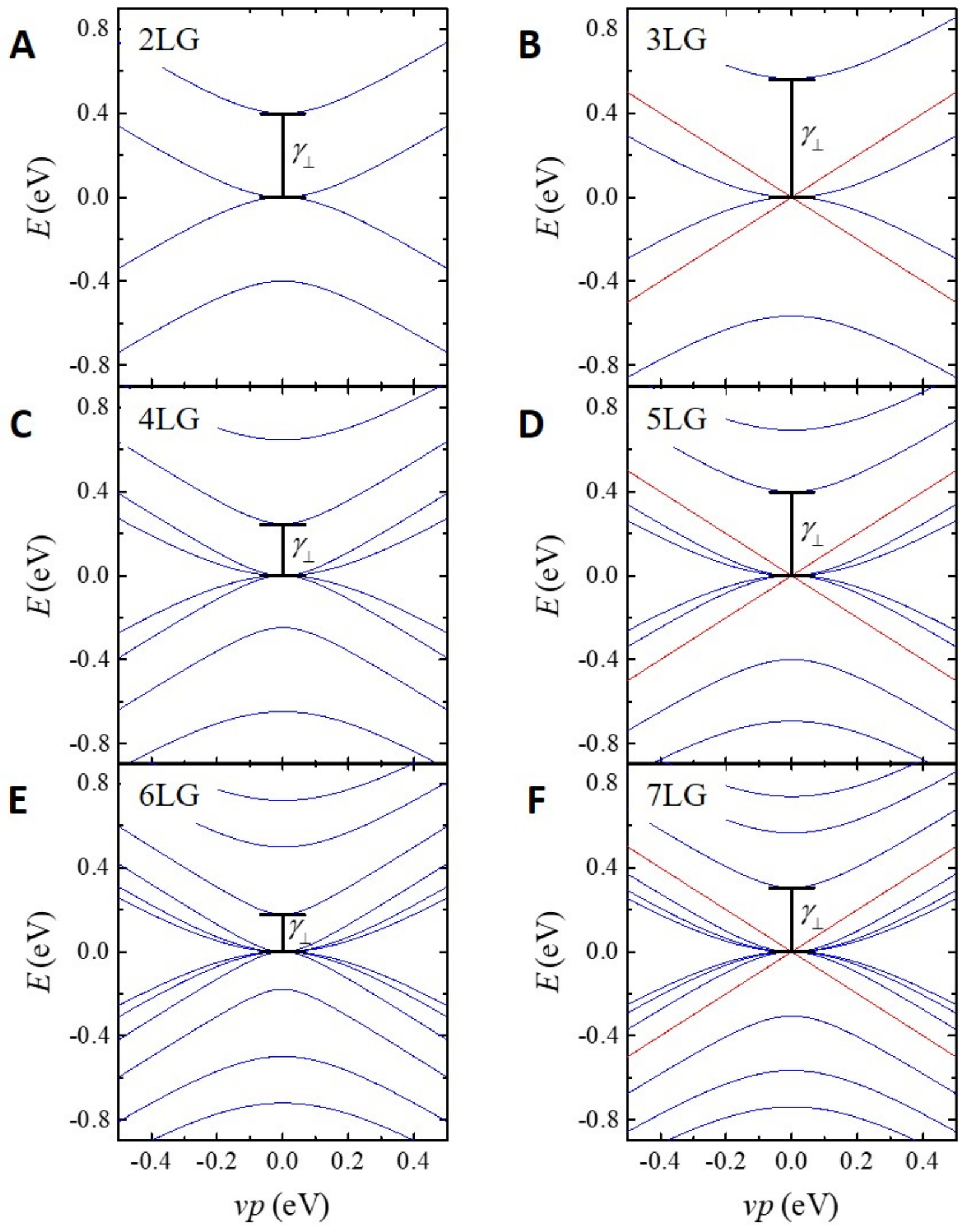

**Fig. S10.**

**Band structure of Bernal-stacked graphene multilayers.** The minimal tight-binding model of Bernal-stacked graphene multilayers assumes that the only non-vanishing hopping integrals are the nearest neighbor intra- and inter-layer terms $\gamma_0$ and $\gamma_1$. Within this approximation, the band structure of Bernal-stacked multilayers consists of one linear Dirac band (only present in odd multilayers) and quadratic bilayer-like bands. Panels (**A**)-(**F**) show the band structure of multilayers of different thickness, from 2LG to 7LG. Even $N = 2M$ multilayers possess $M$ quadratic conduction and valence bands (see (A), (C), (E)), whereas odd $N = 2M+1$ multilayers possess one linear and $M$ quadratic conduction and valence bands (see (B), (D), (F)). All these low-energy bands are degenerate at $E = 0$. Note that each low-energy quadratic band has a corresponding "high-energy band" split-off from zero-energy. The lowest of these "high-energy bands" starts at an energy $\gamma_\perp$. Calculations (see Appendix 1) show that $\gamma_\perp$ decreases with increasing thickness $N$, which prevents –for sufficiently thick multilayers– a well-defined separation of low- and high energy properties. It is for thickness values larger than this scale that the bulk semi-metallic behavior of graphite is eventually recovered.

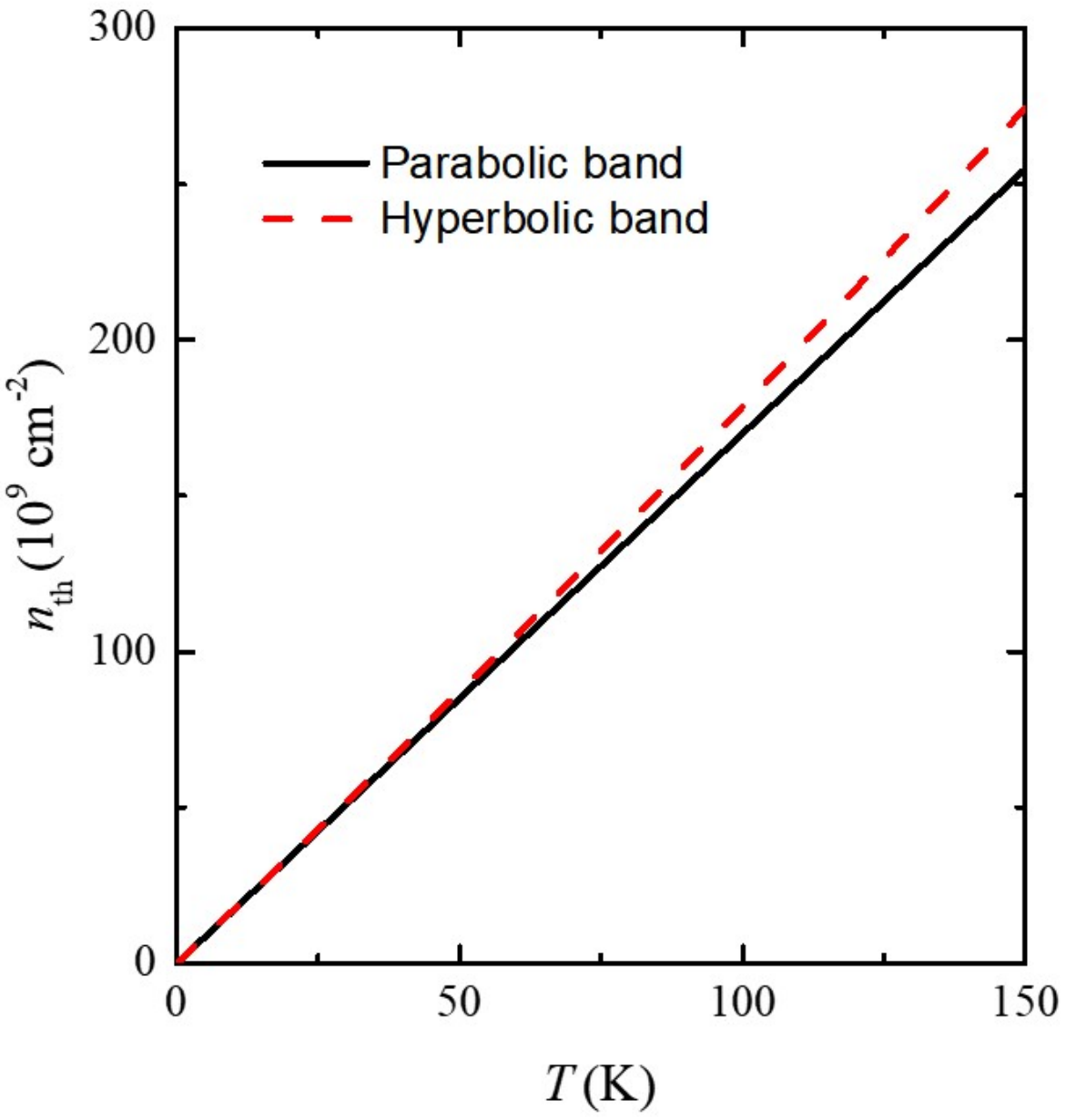

**Fig. S11.**

**Deviation from parabolicity of the low-energy bands in graphene multilayers.** The precise form of band structure of bilayer-like bands obtained from the minimal tight-binding model is hyperbolic. A parabolic dispersion is only valid sufficiently close to zero-energy. Here we compare the density of thermally excited electrons $n_{th}(T)$ present in the conduction band of charge neutral graphene calculated with exact hyperbolic band (dashed red line) and with the approximated parabolic band (continuous black line). The deviation between the two cases is at most a few percent at the highest temperature of our experiments and therefore is essentially negligible. We have checked that the same is true in the range of temperature of our experiments for all multilayers investigated, up to 7LG.